\documentclass[aps,prd,twocolumn,showpacs,superscriptaddress,preprintnumbers,notitlepage,nofootinbib]{revtex4-1}

\usepackage[utf8]{inputenc}
\usepackage{color}
\usepackage{amsmath,amssymb}
\usepackage{amsfonts}
\usepackage{verbatim}
\usepackage{makeidx}
\usepackage{slashed}
\usepackage[vcentermath]{youngtab}
\usepackage{float}
\usepackage{booktabs}
\usepackage{graphicx}
\usepackage{mathrsfs}
\usepackage{bm}
\usepackage{url}
\usepackage{etoolbox}
\apptocmd{\sloppy}{\hbadness 10000\relax}{}{}
\allowdisplaybreaks[3]
\usepackage{color}

\newcommand{\vslash}{{v\hspace{-5.4pt}/}}

\newcommand{\be}{\begin{eqnarray}}
\newcommand{\ee}{\end{eqnarray}}

\newcommand{\del}{\partial}

\newcommand{\bra}{\langle}
\newcommand{\ket}{\rangle}

\newcommand{\calL}{{\cal L}}

\usepackage{ulem}
\definecolor{ao}{rgb}{0.0, 0.5, 0.0}

\definecolor{MO}{rgb}{0.0, 0.0, 0.8}

\begin{document}

\title{$D\bar{D}^\ast$-$\pi J/\psi$ scatterings of coupled channels for $Z_c(3900)$ channel}

\author{Yukihiro~Abe}
\affiliation{Research Center for Nuclear Physics (RCNP), Osaka University, Ibaraki 567-0047, Japan}
\author{Yasuhiro~Yamaguchi}
\affiliation{Department of Physics, Nagoya University, Nagoya 464-8602, Japan}
\affiliation{Kobayashi-Maskawa Institute for the Origin of Particles and the Universe, Nagoya University, Nagoya, 464-8602, Japan}
\affiliation{Department of Physics, Tokyo Metropolitan University, Hachioji 192-0397, Japan}
\author{Atsushi~Hosaka}
\affiliation{Research Center for Nuclear Physics (RCNP), Osaka University, Ibaraki 567-0047, Japan}
\affiliation{Advanced Science Research Center, Japan Atomic Energy Agency, Tokai 319-1195, Japan}
\affiliation{Nishina Center for Accelerator-Based Science, RIKEN, Wako 351-0198, Japan}

\begin{abstract}
We perform 
coupled channel analysis for $D \bar D^*$, $J/\psi \pi$ and related meson pairs for the $Z_c(3900)$ channel
in an effective model of hadrons and quarks.
The model incorporates meson exchange potential such as one pion and $D^{(*)}$ meson exchanges, and quark exchanges.  
It turns out that the meson exchange potential is small, 
while the off-diagonal interactions by the quark exchanges at short distances, 
particularly for transitions between $D\bar D^*$-$J/\psi \pi$ are strong, which plays a main role 
for the scattering amplitudes for the $Z_c(3900)$ channel, in consistent with 
the results of the lattice simulations of the HALQCD group.  
\end{abstract}

\date{\today}
\maketitle

%================================
\section{Introduction}
%================================

While much attention has been paid to exotic hadrons with heavy flavors for decades, 
understanding of their origin is not fully achieved.  
Many exotic hadrons with a clear signal have been observed near and above the threshold region of two or more hadron channels~\cite{Brambilla:2010cs,Chen:2016qju,Guo:2017jvc,Hosaka:2016pey,Yamaguchi:2019vea}.  
A crucial difficulty is in that the interactions between hadrons 
that participate in the relevant reactions for the exotic phenomena
are not well established.  
While remarkable progresses have been made by lattice simulations~\cite{Ishii:2006ec, Aoki:2009ji, Lyu:2022imf, Zhang:2025zaa} and by the analysis of correlation functions~\cite{Fabbietti:2020bfg, Kamiya:2016jqc, Morita:2019rph, Kamiya:2022thy}, their achievements do not cover fully the observed phenomena.   
In general many hadron channels are open above the charm and bottom thresholds 
where exotic phenomena are observed.  
Therefore, coupled channel effects are crucially important with reliable interactions.  
Considering the fact that it took many years even for the nucleon-nucleon interaction 
to be established~\cite{Machleidt:1987hj}, interactions between hadrons for exotic resonances seem extremely difficult.  
In such situations, it is desired to explore an integrated use of various approaches 
including effective models, experimental data and lattice simulations.

Motivated partly by the result of the lattice simulations of the HALQCD collaboration~\cite{Ikeda:2016zwx,Ikeda:2017mee}, 
in this paper 
we study the interactions for the channel of $Z_c(3900)$ in an effective models for hadrons and quarks.  
The $Z_c(3900)$ is a candidate of an exotic particle containing $\bar c c$ and $u \bar d$, and hence
it is a genuine exotic tetraquark, if it exists.  
The signal was first observed by BESIII~\cite{BESIII:2013ris}, and then also confirmed by
Belle~\cite{Belle:2013yex} and CLEO~\cite{Xiao:2013iha}, 
where they have analyzed the invariant mass spectrum of $\pi^{\pm} J/\psi$
in the process $e^+ e^- \to \pi^+ \pi^- J/\psi$.  
Subsequently, BESIII observed a signal also in the open charm channel 
of $D \bar D^*$~\cite{BESIII:2013qmu,BESIII:2015pqw}.  
Although more than ten years have passed, its nature is not yet well understood.  
Then a lattice simulation was performed for coupled-channels including 
 $D\bar{D}^\ast$-$\pi J/\psi$-$\rho \eta_c$ corresponding to $Z_c(3900)$
 at the non physical pion mass $m_\pi=410-700$ MeV~\cite{Ikeda:2016zwx,Ikeda:2017mee}.
The finding of the lattice simulation is rather nontrivial.
It indicates only weak interactions for the diagonal open-charm channels of  $D\bar D^*$, while the interaction for the off-diagonal ones such as $D\bar D^*$-$J/\psi \pi$ is 
significantly large.  
Using the obtained interaction, they have analyzed coupled-channel scattering amplitudes, 
and found only a virtual state pole for $Z_c(3900)$ with the resulting line shape of the observed mass distribution qualitatively consistent with the data.

Recently, rather inclusive amplitude analyses have been performed 
with coupled channels~\cite{Yu:2024sqv,Nakamura:2023obk}.   
In Ref.~\cite{Yu:2024sqv}, a model of three coupled channels $D \bar D^{*}, J/\psi \pi$ and 
$\rho \eta_c$
was employed with the inclusion of triangle singularities.  
By using the data of Refs.~\cite{BESIII:2013qmu,BESIII:2015pqw}, 
they  performed an invariant mass analysis of relevant two-body channels,  
and  concluded that the signal of $Z_c(3900)$ was likely to be a cusp.  
In Ref.~\cite{Nakamura:2023obk}, they have performed a global analysis 
with more than ten relevant channels for the $e^+ e^- \to c \bar c$ process.  
The state-of-arts analysis implied that the $Z_c(3900)$ was a virtual state 
whose pole was located $\sim 40$ MeV below the $D \bar D^*$ threshold.  
In both studies, the importance of interactions was emphasized.

In the present paper we employ a coupled channel model with
$D^{(*)} \bar{D}^{(*)}$-$\pi \psi^{(\prime)}$-$\rho \eta_c$  channels with interactions 
in an effective model of meson exchanges~\cite{Hosaka:2016ypm,Yamaguchi:2019vea} and quark exchanges~\cite{Barnes:1991em,Swanson:1992ec}. 
Here  $D^{(*)}$ indicates $D$ or $D^*$, and  $\psi^{(\prime)}$ includes $J/\psi(1S)$ and $\psi(2S)$, 
focusing our study on the channel of $Z_c(3900)$, namely the spin and parity 
$J^{PC} = 1^{+-}$ and isospin $I = 1$.  
The $\pi, \rho, \omega$ exchanges are considered for the diagonal open charm channels
$D^{(*)} \bar{D}^{(*)}$-$D^{(*)} \bar{D}^{(*)}$, 
while $D^{(*)}$ exchange is considered for the off-diagonal 
$D^{(*)} \bar{D}^{(*)}$-$\pi J/\psi$(-$\rho \eta_c$) interaction.  
In contrast, the quark exchange contributes only to the off-diagonal interaction 
between open-charm and hidden-charm channels.  
The resulting interaction is then compared with the lattice results.

One of our motivations in this work is to see whether we can understand the lattice results 
in terms of effective models.
We do this for a particular channel of $Z_c$.  
However, if we achieve a reasonable agreement with the lattice results, we may apply 
the similar idea based on effective models to other systems.

This paper is organized as follows.
After a brief summary for the nominal $Z_c$ state in 
section~\ref{Sec:Zc-state}, in section~\ref{Sec:D-exchange}
we derive the meson exchange potential for the one pion 
and $D^{(*)}$ meson exchanges.  
In both cases, we show that the interaction is reasonably approximated 
in the form of a static potential.  
Section \ref{Sec:q-exchange}
derives the quark exchange potential with the one-gluon exchange.  
In 
section~\ref{Sec:Results}
we show numerical results for various components 
of the potential, 
and present scattering amplitudes by solving 
coupled channel equations with relevant two-meson coupled channels.  
In that section, we compare our potential with the lattice results 
at the quark masses that were used by the lattice simulations.  
The final 
section~\ref{Sec:Summary}
is devoted discussions and summary.  

%================================
 \section{\label{Sec:Zc-state} $Z_c$ state}
%================================

To start with, let us briefly look at the relevant coupled channels 
of hadrons for $Z_c$.
Its quantum numbers of spin, parity and charge conjugation are
\be
J^{PC} = 1^{+-}
\ee
They are formed by various combinations of the open charm and hidden charm mesons, 
$D^{(*)} \bar D^{(*)}$ and $\pi J/\psi, \ \pi \psi^\prime, \ \rho \eta_c$.  
The hidden charm combinations, $\pi J/\psi, \ \pi \psi^\prime, \ \rho \eta_c$, satisfy by themselves the condition $J^{PC} = 1^{+-}$
together with two-meson orbital wave function of  angular momentum ($L = 0, 2$). 
For $D^{(*)} \bar D^{(*)}$, suitable combinations are formed 
to satisfy spin 1 and negative charge conjugation.  
They are 
\be
\frac{1}{\sqrt{2}}(D \bar D^* + D^* \bar D) \psi_L(x), \ \ \
D^* \bar D^* \chi_1 \psi_L(x)
\label{CCeigenstate}
\ee
where $\psi_L$ is the orbital wave function of an even angular momentum $L$, 
and $\chi_1$ is the spin 1 wave function.  
For the $D \bar D^*$ term the spin 1 wave function is trivially formed, 
while for $D^* \bar D^*$ the two spin 1 states are formed 
in anti-symmetric manner.  

%================================
 \section{\label{Sec:D-exchange} Meson exchange interaction}
%================================

\begin{figure}[h]
  \begin{center}
\includegraphics[width=0.9\linewidth,clip]{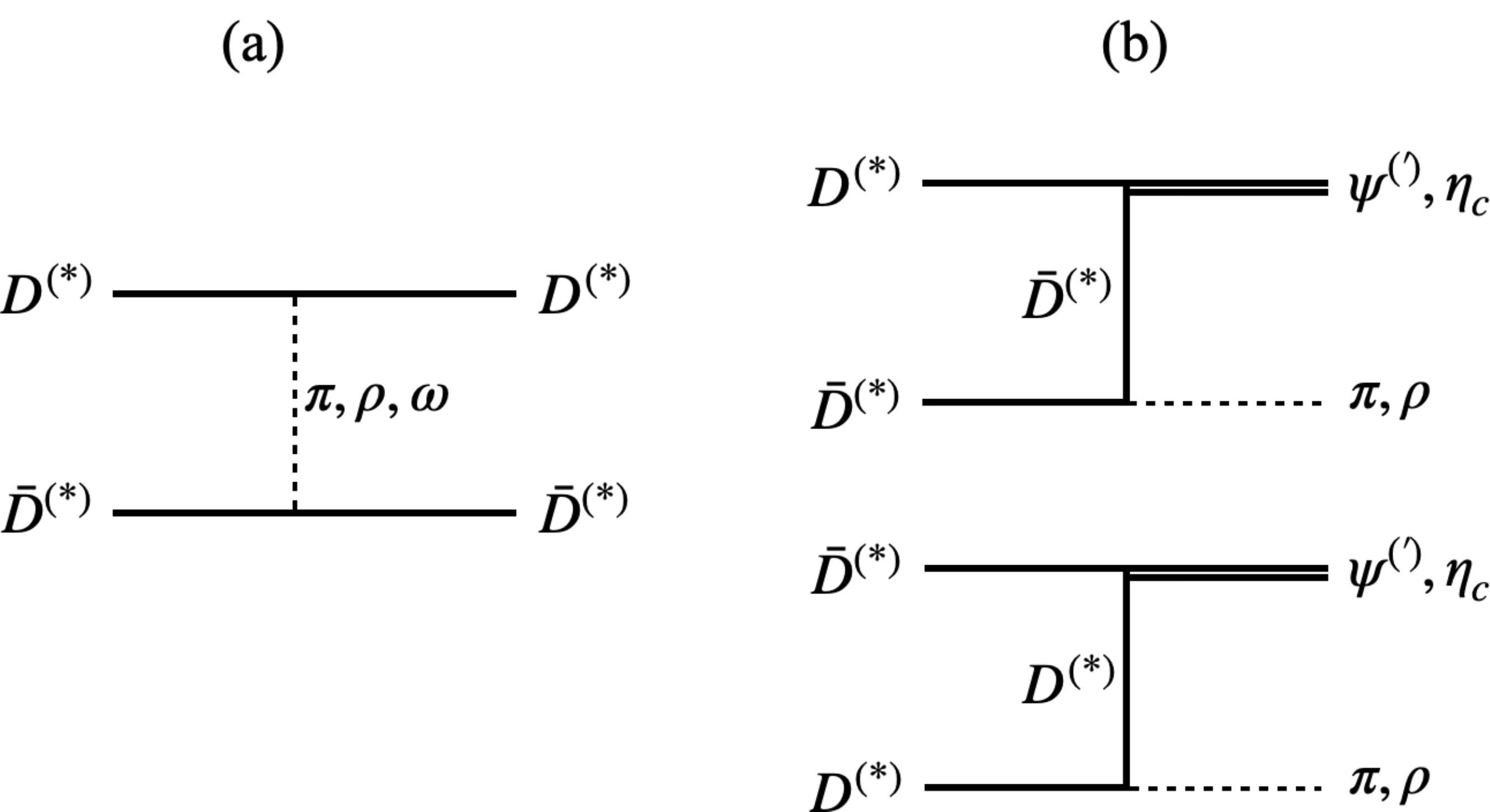}
   \caption{\label{fig_meson_exchange} Meson exchange diagrams.}
  \end{center}  
 \end{figure}

In this section we show the meson exchange interaction for a coupled channel system of $D^{(*)} \bar{D}^{(*)}$-$\pi \psi^{(\prime)}$-$\rho \eta_c$
 as shown in Fig.~\ref{fig_meson_exchange}.
 The necessary Lagrangians of Yukawa vertices are given 
by~\cite{Casalbuoni:1996pg,Yamaguchi:2019vea,Miyake:2025ktz,Wang:2015xsa}
 \begin{align}
  {\cal L}_{\pi H H}
    = & 
  -\frac{g_\pi}{2f_\pi} {\rm Tr}\left[H_1
  \gamma_\mu\gamma_5 \partial^{\mu}\pi\bar{H}_1\right] \notag\\
  & 
    -\frac{g_\pi}{2f_\pi} {\rm Tr}\left[\bar{H}_2
  \gamma_\mu\gamma_5 \partial^{\mu}\pi{H}_2\right], 
  \ \ \  
  (\pi \equiv  \bm{\tau} \cdot \bm{\pi})
  \label{L_piHH} \\
{\cal L}_{V H H }= & 
    -i\beta {\rm Tr}\left[H_1 v^\mu \hat{V}_\mu \bar H_1 \right]
  + i \lambda {\rm Tr}\left[H_1 \sigma^{\mu \nu} F_{\mu \nu} \bar H_1\right] \nonumber \\
  & +i\beta {\rm Tr}\left[\bar{H}_2 v^\mu V_\mu H_2 \right]
  + i \lambda {\rm Tr}\left[\bar{H}_2 \sigma^{\mu \nu} F_{\mu \nu} H_2\right], \nonumber \\ 
\hat{V}_\mu \equiv & 
\frac{ig_V}{2} V_\mu
 = \frac{ig_V}{2} \left(\bm{\tau} \cdot \bm{\rho}_\mu + \omega_\mu \right), \nonumber \\  
F_{\mu \nu} = & 
\del_\mu \hat{V}_\nu - \del_\nu \hat{V}_\mu, \ \ \  
\label{L_VHH} \\
 {\cal L}_{\psi \bar H_1 \bar H_2}
 = & 
 g^\prime {\rm Tr}\left[{\cal J}{\bar H}_2\overset{\leftrightarrow}{\partial}_\mu\gamma^\mu{\bar H}_1\right] 
  + {\rm H.c.},
 \end{align}
where $i = 1, 2$,  and $H_1$, $H_2$ and $  {\cal J}$ are for the annihilation 
of heavy quark multiplets, $(D, D^*)$, $(\bar D, \bar D^*)$ and ($\psi, \eta_c$) 
mesons, respectively, 
 \begin{align}
  H_1&=\frac{1+\vslash}{2}\left[D^{\ast\mu}\gamma_\mu+iD\gamma_5\right], 
  \nonumber \\
  H_2&=\left[\bar{D}^{\ast\mu}\gamma_\mu+i\bar{D}\gamma_5\right]\frac{1-\vslash}{2}, 
  \nonumber \\
  {\cal J}&=\frac{1+\vslash}{2}\left[\psi^\mu\gamma_\mu+i\eta_c\gamma_5\right]\frac{1-\vslash}{2} .
  \nonumber
 \end{align}
We need the two fields $H_{1,2}$, because in the heavy quark limit
$D^{(*)}$ and $\bar D^{(*)}$ are regarded as different particles, and their creation and
annihilation are suppressed.  
For the phenomena near the $D^{(*)}\bar D^{(*)}$ threshold region that we are interested in 
we can treat the scattering of heavy particles $D^{(*)}, J/\psi$ in a non-relativistic manner which allows to set 
$v_\mu = (1, 0, 0, 0)$.  
We employ the coupling constants as 
$g_\pi=0.59$, 
$\beta = 0.9$, $\lambda = 0.56$ GeV$^{-1}$, $g_V\sim 5.8$, 
$g^\prime=4/\sqrt{m_\psi m^2_D}$, and the pion decay constant $f_\pi=93$ MeV~\cite{Casalbuoni:1996pg,Yamaguchi:2019vea,Miyake:2025ktz,Wang:2015xsa}.

%-------------------------------
%-------------------------------

For practical purposes, it is convenient to prepare the meson exchange interaction 
in the form of a static potential in the coordinate space.  
There are two cases that are qualitatively different 
as shown in Fig.~\ref{fig_meson_exchange} (a) and (b).

The case (a) is the light meson 
($\pi$, $\rho$ and $\omega$)
exchange between the $D^{(*)}\bar D^{(*)}$ channels.  
Here we demonstrate the pion exchange potential, 
while other vector meson exchanges are summarized in 
 Appendix~\ref{appendix:Vectorexchange}.
In the heavy quark limit where the masses of $D$ and $D^*$ are degenerate, 
the energy transfer is neglected and the standard static potential is obtained.
When the mass difference between $D$ and $D^*$ is taken into account, 
the mass of the exchanged pion 
should be replaced by an effective one $\mu$~\cite{Yamaguchi:2019vea}.  
Therefore, the one pion exchange potential for 
$D^{(*)}\bar{D}^{(*)} \to D^{(*)}\bar{D}^{(*)}$ 
is given by 
\be
& & 
V_{D^{(*)}\bar{D}^{(*)} \to D^{(*)}\bar{D}^{(*)}}^\pi (\bm r)
=
   -\frac{1}{3}\left(\frac{g_\pi}{2f_\pi}\right)^2
   \nonumber \\
 & &\ \ \ \  \times
   \left[\bm{S}_1\cdot\bm{S}_2C(r;\mu)+S_{12}(\hat{r})T(r;\mu)\right]
     \bm{\tau}_1\cdot\bm{\tau}_2, 
     \label{V_pi}
\ee
where the spin operator $\bm{S}_{i = 1,2}$ at $DD^*\pi$ vertex 
is the polarization vector of $D^*$, $\bm \epsilon$, 
and at $D^*D^*\pi$ vertex the $3 \times 3$ spin one matrix.
The tensor operators are then formed as 
$S_{12}(\hat {\bm r}) = 
3 \bm S_1 \cdot \bm \hat {\bm r}\  \bm S_2 \cdot \hat {\bm r} - \bm S_1 \cdot \bm S_2$, etc.

The radial functions are 
\be
C(r; \mu) &=&
\int \frac{d^3\bm q}{(2\pi)^3} \frac{\mu^2}{\bm q^2 + \mu^2} e^{i \bm q \cdot \bm r} F(\Lambda, \bm q), 
\nonumber \\
S_{12}(\hat {\bm r}) T(r; \mu) &=&
\int \frac{d^3\bm q}{(2\pi)^3} \frac{-\bm q^2}{\bm q^2 + \mu^2} e^{i \bm q \cdot \bm r} 
S_{12}(\hat {\bm q})  F(\Lambda, \bm q)
\nonumber \\
F(\Lambda, \bm q) &=& \left( \frac{\Lambda^2 - m^2}{\Lambda^2 - q^2} \right)^2
\ee
We also note that for the spin-spin interaction $C(r; \mu)$,
the constant term (delta function in coordinate space) has been removed, and 
instead, the form factor $F(\Lambda, \bm q)$ has been introduced.  
The cutoff parameter $\Lambda$ is determined by the following formula
as inspired by the relation to the size of hadrons,  
\be
\Lambda = \Lambda_N \times 1.35, \ \ \ r_N/r_D = 1.35
\ee
Here $\Lambda_N = 836 $ MeV is employed to reproduce the deuteron properties 
such as binding energy, the scattering length and effective range of the nucleon scattering.  
The factor 1.35 is the ratio of the sizes of the nucleon and the $D^{(*)}$ meson 
with the physical basis that the form factor reflects the spatial sizes of the scattering particles.  
This is a prescription inspired by the one pion exchange potential for $D^{(*)}$ mesons, 
and is also applied to the $\rho$ meson exchanges.  
Details are given in 
Refs.~\cite{Yamaguchi:2019vea,Yamaguchi:2011xb,Ohkoda:2011vj,Hosaka:2016ypm,Yamaguchi:2016ote}.

The case (b) is for the transitions between $D^{(*)}\bar D^{(*)}$ and $J/\psi\pi$ 
as in Fig.~\ref{fig_meson_exchange} (b), where 
the relevant diagrams which conserve the number of heavy quarks are shown in chronological order.
In that process the fact that a large energy transfer should be taken into account carefully.
For this let us  consider the following scattering process in the center of mass system, 
\be
& & D^{(*)}(E_D(\bm p), \bm p) +  
\bar{D}^{(*)}(E_{\bar{D}}(\bm p), -\bm p) 
\nonumber \\
& & \to \pi(E_\pi(\bm p^\prime), \bm p^\prime) + J/\psi(E_{J/\psi}(\bm p^\prime), -\bm p^\prime) 
\ee
where inside braces are the energies and momenta of the particles.  
We assume that the momentum $\bm p$ is small near the threshold region of 
$D^{(*)}\bar D^{(*)}$, and consider the heavy quark mass limit 
where the masses of $D$ and $D^*$ mesons are degenerate.  
In this limit, we may replace the energies of heavy particles by their masses, e.g., 
\be
E_D(\bm p) \to m_D, \ \ \  E_{J/\psi} \to m_{J/\psi}
\ee
while the energy of the pion is kept as
\be
E_\pi(\bm p^\prime) = \sqrt{m_\pi^2 + \bm p^{\prime 2}} \sim 2m_D - m_{J/\psi}.
\label{Epi}
\ee
Having these kinematic approximations, we can compute the transition amplitudes.

As shown in 
Appendix~\ref{appendix:Dexchange},
in the limit of heavy mass of $D^{(*)}$, 
the amplitude is a constant in momentum space
except for the tensor structure.  
Specifically, $D$ meson exchange for $D \bar D^* \to J/\psi \pi$, is given as
\be
\tilde V_{D \bar D^* \to J/\psi \pi}^D  = 
- \frac{g_\pi g^\prime }{\sqrt{2}f_\pi E_\pi^{3/2}} 
\bm S_1 \cdot \bm q  \bm S_2 \cdot \bm q
\ee

As in the case of the pion exchange, we multiply a form factor due to the finite structure of scattering mesons.  
The momentum transfer $\bm q$ is then transformed into the derivative acting on it.  
The form factor is introduced as the Fourier transform of the overlap 
of the incoming and outgoing mesons at the vertex, 
\be
F(\bm q) = \int d^3x e^{i\bm q \cdot \bm x} 
\psi^\dagger_{\rm out}(\bm x) \psi_{\rm in}(\bm x)
\ee
where $\psi_{\rm in, out}(\bm x)$ are the internal wave functions of the mesons, and 
$\bm q$ is the momentum of the exchanged meson.  
Employing the harmonic oscillator wave functions as in the standard quark model, 
for the transitions between the ground state mesons we have 
\be
F(\bm q) = N e^{- \bm q^2/\Lambda^2}
\ee
where $N$ is the normalization factor and 
the cutoff parameter $\Lambda$ is determined by the sizes of the mesons.  
When the two mesons have the same size, $N = 1$, 
while if they are different, $N < 1$ due to an incomplete overlap.  
The form factor is attached at each vertex of meson exchange diagrams, which is labeled by 
an index 1 or 2.  

Hence we arrive at the expression
\be
\tilde V(\bm q)_{D \bar D^* \to J/\psi \pi}
=
- G
\bm S_1 \cdot \bm q  \bm S_2 \cdot \bm q N_1 N_2 e^{- \bm q^2/\Lambda_{12}^2}
, \ 
\label{eq_V(q)}
\ee
where 
\be
G = - \frac{g_\pi g^\prime }{\sqrt{2}f_\pi E_\pi^{3/2}} , 
\ \ \
\frac{1}{\Lambda_{12}^2} = \frac{1}{\Lambda_{1}^2} + \frac{1}{\Lambda_{2}^2}
\ee
Performing the Fourier transform, we find
\be
V_{D \bar D^* \to J/\psi \pi}(\bm r) = G N_1 N_2
\left[
\bm S_1 \cdot \bm S_2 C(r) + S_{12}(\hat r) T(r)
\right]
\nonumber \\
\label{VDD*Jpsipi}
\ee
The radial functions $C(r)$ and $T(r)$ are given as
\be
C(r) &=& 
- \frac{3\Lambda_{12}^5}{16\pi^{3/2}} 
\left( 1-\frac{\Lambda_{12}^2 r^2}{6} \right)
e^{-\Lambda_{12}^2 r^2/4}
\nonumber \\
T(r) &=& \frac{\Lambda_{12}^7}{32 \pi^{3/2}} e^{-\Lambda_{12}^2 r^2/4}
\ee
Potentials for other channels are given in 
Appendix~\ref{appendix:Dexchange}.

\begin{figure}[h]
  \begin{center}
\includegraphics[width=0.9\linewidth,clip]{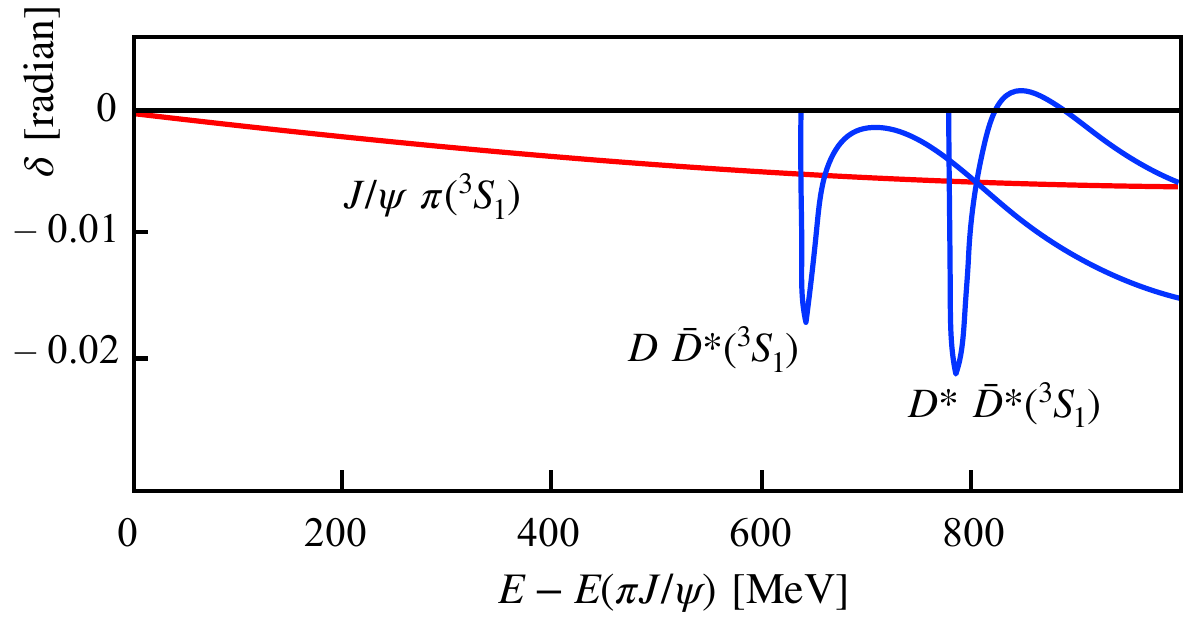}
   \caption{\label{fig_phaseshift_meson} Phase shifts obtained by using the meson exchange potential for various channels as shown in the panel.}
  \end{center}  
 \end{figure}

Using the obtained meson-exchange interaction, we have computed 
phase shifts.
It turns out that the resulting phase shifts are tiny, and so are the amplitudes.  
In Fig.~\ref{fig_phaseshift_meson}, phase shifts for the S-wave scattering of $\psi$-$\pi$ and of $D$-$\bar D^*$ 
for $I = 1$ and 
$J^{PC} = 1^{+-}$ 
channels.
The  small $J/\psi\ \pi$ phase shift is expected because the interaction is from 
the coupled channel of $D \bar D^*$ and $D^* \bar D^*$ with small coupling strengths.  
As for the $D^{(*)} \bar D^{(*)}$ channel, the one pion exchange potential has the effective mass as an overall factor, $\mu < m_\pi$.  
In addition,  the isospin factor 
$\bm \tau_1 \cdot \bm \tau_2$ which is, for the isospin 1 channel, 
1/3 of the isospin 0 channel.  
Hence, the one pion exchange potential is small.  
The $\rho$ meson exchange is also small due to the form factor 
$(\Lambda^2 - m_\rho^2)/(\Lambda^2 - q^2)$.

As we have verified that meson exchange process for the $Z_c$ channel 
turned out to be very small and is far from what is expected in the experimental data
and from the lattice result, 
we shall investigate quark exchange processes in the next section.

%-------------------------------
%-------------------------------

%================================
 \section{\label{Sec:q-exchange} Quark exchange interaction}
%================================

The quark exchange process for two-meson scattering 
has been considered for example
by Barns and Swanson~\cite{Barnes:1991em,Swanson:1992ec}.
Quark exchanges between two mesons occur via interactions, 
for which we consider one-gluon exchange as 
often used in the non-relativistic quark model.
It consists of the  Coulomb, linear confinement and spin-spin terms
with color factors in common, 
\begin{align}
v = \sum_{i<j}
\left( v^{\rm Coul}_{ij}+v_{ij}^{\rm conf} +v_{ij}^{ ss} \right)
\frac{\lambda^a_i}{2} \frac{\lambda^a_j}{2} 
\label{eq_V_quark}
\end{align}
where
\begin{align}
v^{\rm Coul}_{ij} &=\frac{\alpha_C}{r_{ij}}, \ \ \ 
v^{\rm conf}_{ij} = - \frac{3}{4}br_{ij}, \\
v_{ij}^{ ss} &= 
-\frac{8\pi \alpha_h}{3m_i m_j}\left(\frac{\sigma^3}{\pi^{3/2}}e^{-\sigma^2 r^2_{ij}}\right) \bm{S}_i\cdot \bm{S}_j
 \end{align}
In these equations, the indices $i, j$ labels the interacting quarks, $r_{ij}$ their inter distance, 
$\lambda^a$ ($ a = 1, \cdots 8)$ are Gell-Mann matrices for colors and 
$\bm{S}_j = \bm{\sigma}_j/2$  spin matrices.
The parameters $\alpha_c, b, \alpha_h, \sigma$ are taken from Ref.~\cite{Silvestre-Brac:1996myf}

For the meson wave functions that we need for the quark exchanges, we approximate them by one range (term) Gaussian function which minimizes the Hamiltonian with the potential (\ref{eq_V_quark}) and the kinetic term.  
Then the light quark mass is tuned to reproduce the mass splitting of the $\pi$ and $\rho$ mesons in the lattice calculation (Case I in Ref.~\cite{Ikeda:2017mee}), while the charm quark mass is set at the one of Ref.~\cite{Silvestre-Brac:1996myf},
\begin{align}
m_q = 361\ {\rm MeV}, \ \ \ 
m_c = 1840\  {\rm MeV}
\end{align}
The resulting meson masses are summarized in Table~\ref{table_quarkmasses}.  
In the present numerical calculations we employ this parameter set.  
In addition to this parameter set, we have performed calculations by using other sets within a reasonable range, which turns out not to change the essential features that we will show in the next sections.  

\begin{table}[htp]
\caption{Meson masses used in this work and those in the lattice calculation~\cite{Ikeda:2017mee} in units of MeV. }
\begin{center}
\begin{tabular}{c c c}
\hline
Meson & This work & Lattice \\
\hline
$m_\pi$ &  313   & 411  \\
$m_\rho$ & 798  &  896 \\
$m_{J/\psi}$ &  3121   &  3097 \\
$m_{\eta_c}$ &  3032  &  2988 \\
$m_D$ &  1892  &  1903 \\
$m_{D^*}$ &  2028  & 2056 \\
\hline
\end{tabular}
\end{center}
\label{table_quarkmasses}
\end{table}%

\begin{figure}[h]
  \begin{center}
\includegraphics[width=1\linewidth,clip]{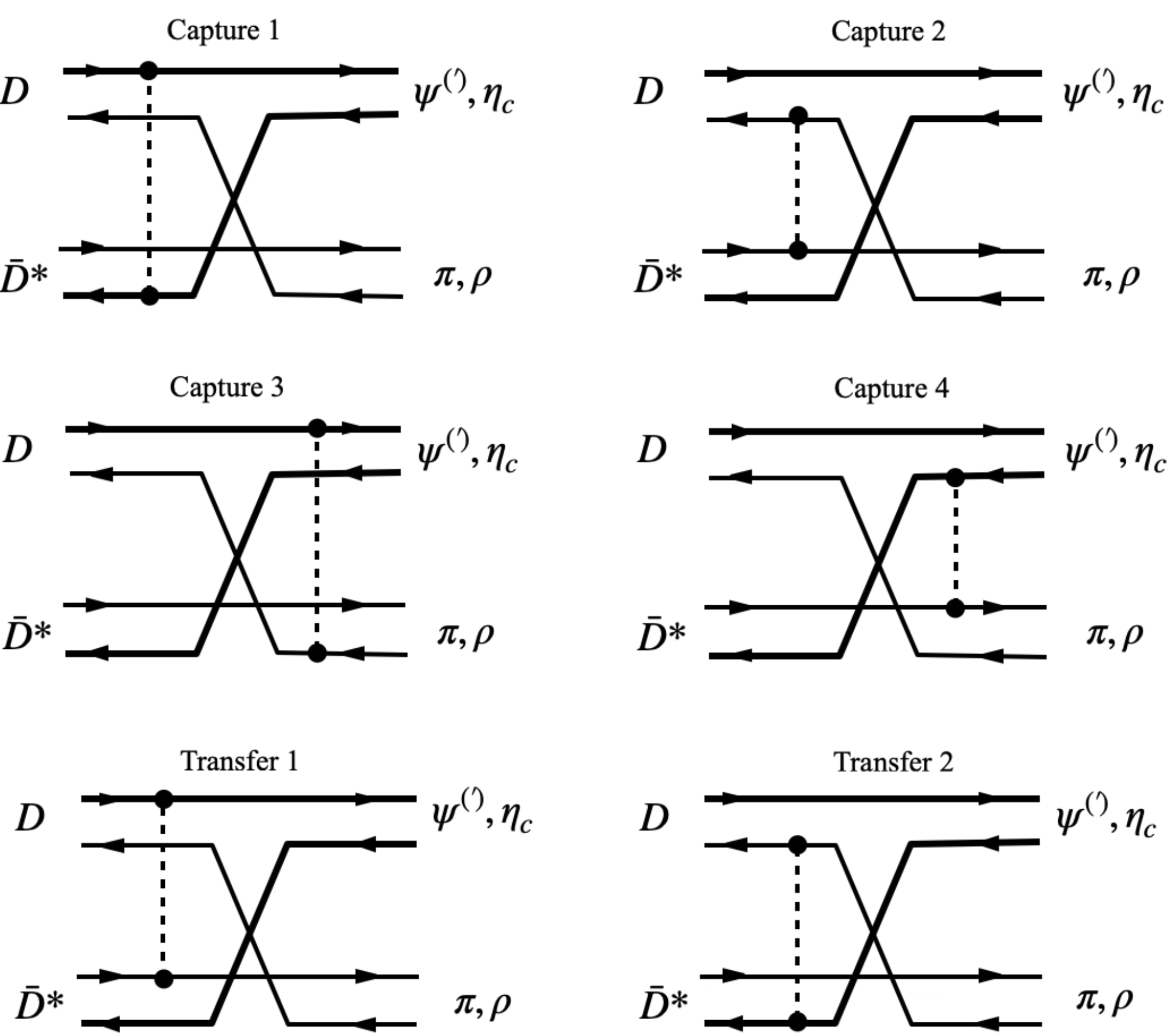}
   \caption{\label{fig_q_exchange_full} Six types of quark exchange diagrams.}
  \end{center}  
 \end{figure}

The quark exchange interaction is effective only for the transitions between 
the open and hidden charm channels via heavy (charm) quark exchange 
such as  $D \bar D^* \to \pi J/\psi$, 
while it is ineffective for the diagonal channel $D \bar D^* \to D \bar D^*$ because 
of the color singlet nature of hadrons.  
The relevant diagrams are shown in Fig.~\ref{fig_q_exchange_full}, where 
the classification of ``Capture" or ``Transfer" follows Refs.~\cite{Barnes:1991em,Swanson:1992ec}.

There is a subtlety in the six processes.  
Because the initial and final states are different, 
the equivalence between the post and prior form, or in general, 
the symmetry of the T-matrix  is guaranteed 
either when including all the six processes or when including the two transfer processes.  
For the capture processes, the interactions shown in Fig.~\ref{fig_q_exchange_full}
could be interpreted as the one for the formation of the meson states.  
In fact, the interaction (\ref{eq_V_quark}) 
that we use for the present study for hadron interaction 
is used for the formation of the meson wave functions.  
(Actually in the present study they are approximated by the Gaussian functions as shown in appendices.)
If this would be the case, the relevant process is the transfer ones.  
Having this discussion, we will investigate 
contributions from the Capture, Transfer and their sum, respectively in the next section.

%-------------------------------
%-------------------------------

\begin{figure}[h]
  \begin{center}
\includegraphics[width=0.6\linewidth,clip]{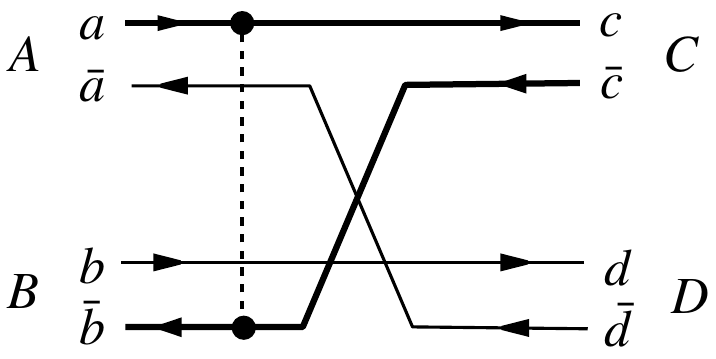}
\caption{
\label{fig_q_exchange_ex} 
   Definition of  various momenta for the process 
Capture 1 in Fig.~\ref{fig_q_exchange_full}.
       }      
  \end{center}  
 \end{figure}

To show concrete calculations 
we consider an example for a two-meson scattering as shown 
in Fig.~\ref{fig_q_exchange_ex}
\be
\label{eq_ABtoCD}
A(a \bar a) + B(b \bar b) \to C(c \bar c) + D(d \bar d)
\ee
where momentum variables for particles $A, B, \cdots$  are 
denoted by bold letters 
such as $\bm A, \bm B, \cdots$, 
and those of the constituent quarks and antiquarks $a, \bar a$ 
by the small and bold letters, 
$\bm a, \bar {\bm a}, \bm b, \bar {\bm b}, \cdots$.

%-------------------------------
%-------------------------------

For the capture 1 the quarks $a$ and $\bar b$ interact via $v_{ij = a \bar b}$, 
the Born amplitude is computed by 
\begin{align}
&  \langle C,D|v_{a \bar b}|A,B \rangle \\
& =  \,
 \left(-\frac{4}{9}\right) 
 I_{spin}^{(c1)}
 \frac{-1}{(2\pi)^3} 
 \delta^{(3)}(\bm{C}+\bm{D}-\bm{A}-\bm{B})
 \notag\\
 & 
\times 
 \int d^3a \,d^3c  
 \, 
 \Phi_{C}^\ast(\bm{c_r}^{(c1)})\Phi_{D}^\ast(\bm{d_r}^{(c1)})
v_{a \bar b}(|\bm{c}-\bm{a}|)  \notag\\
 & \times
 \Phi_{A}(\bm{a_r}^{(c1)})\Phi_{B}(\bm{b_r}^{(c1)}) , 
 \label{eq:BornAmp_1}\\
 & \equiv (2\pi)^3 \delta^{(3)}(\bm{C}+\bm{D}-\bm{A}-\bm{B}) h^{(c1)}_{fi} ,
\end{align}
where 
\begin{align}
 \bm{a_r}^{(c1)} &
 = \bm{a} - \frac{m_a}{m_a+\bar{m}_a}\bm{A} ,
 \\
 %%%
 \bm{b_r}^{(c1)} &
 = \bm{a} - \bm{C} - \frac{\bar{m}_b}{m_b+\bar{m}_b}\bm{A} ,
 \\
 %%%
  \bm{c_r}^{(c1)} &
 = \bm{c} - \frac{m_c}{m_c+\bar{m}_c}\bm{C} ,
 \\
 %%%
 \bm{d_r}^{(c1)} &
 = \bm{a} - \bm{A} - \frac{\bar{m}_d}{m_d+\bar{m}_d}\bm{C} . 
\end{align}
In these expressions, $\Phi_{\alpha}$ are the internal wave functions of the mesons $\alpha = A, B, C, D$ in momentum space, and the spin factors $ I_{spin}^{(i)}$ are shown in 
Appendix~\ref{appendix:Spinfactor}.

Other diagrams are computed by replacing the momenta suitably 
depending on the process shown in Fig.~\ref{fig_q_exchange_ex}.  
Complete expressions are presented in 
Appendix~\ref{appendix:BornAmp}.
The sum over all diagrams in Fig.~\ref{fig_q_exchange_full} is the  Born amplitude
that we identify with the (transition) potential.   

In general the amplitudes depend on the two momentum variables, 
$
\bm p_i = \bm A - \bm B, \  \bm p_f = \bm C - \bm D, 
$
or equivalently, 
$
\bm q = \bm p_f - \bm p_i, \  \bm P = \frac{1}{2} (\bm p_i + \bm p_f), 
$
\be
V(\bm p_f, \bm p_i) = V(\bm q, \bm P) \equiv \sum_{ij} \bra C,D| v_{ij}  |A,B\ket .
\ee
We note that the sum over the pair of quarks $ij$ completes the sum over the diagrams in Fig.~\ref{fig_q_exchange_full}, because each diagram there determines the pair of quarks 
between which one gluon is exchanged. 
Introducing the coordinates $\bm r_i$ and $\bm r_f$  
which are conjugate to $\bm p_i$ and $\bm p_f$, 
we find 
\be
\bra \bm r_f | V | \bm r_i\ket 
&=& \int \frac{d^3p_f}{(2\pi)^3} \frac{d^3 p_i}{(2\pi)^3}  \ e^{i\bm p_f \cdot \bm r_f - i \bm p_i \cdot \bm r_i} 
V(\bm p_f, \bm p_i)
\nonumber \\
&=& \int \frac{d^3P}{(2\pi)^3}  \frac{d^3 q}{(2\pi)^3}  \ e^{\frac{i}{2} \bm P \cdot \bm \rho - i \bm q \cdot \bm r}
V(\bm q, \bm P)
\ee
where we have defined 
\be
\bm r = \frac{1}{2}(\bm r_i + \bm r_f), \ \ \ \bm \rho = \bm r_f - \bm r_i
\ee
The $\bm P$-dependence leads to non-locality as it shows 
the dependence on the difference in the initial and final coordinates, 
$\bm \rho = \bm r_f - \bm r_i$.  
In the present work, we assume that such non-locality is not very strong, 
$V(\bm q, \bm P) \sim V(\bm q)$, and hence $\bm r_f \sim \bm r_i$.   
In this way we find a local potential 
(by removing the delta function $(2\pi)^3\delta(\bm r_f - \bm r_i)$)
\be
V(\bm r) = \int \frac{d^3 q}{(2\pi)^3}  e^{-i \bm q \cdot \bm r} V(\bm q)
\label{eq_localV}
\ee
More complete derivation is shown in 
Appendix~\ref{appendix:Fouriertrans}.
In the next section 
this will be compared with the results of the lattice calculation~\cite{Ikeda:2016zwx,Ikeda:2017mee}.  

%================================
 \section{\label{Sec:Results} Results }
%================================

%---------------------------------
\subsection{\label{Scattering} Scattering equations in coupled channels}
%---------------------------------

We solve the following coupled channel equation, 
\be
(K + V) \psi = E\psi
\label{eq_Schro_eq}
\ee
where $K$ is the kinetic term.  
The relevant channels are 
\be
J/\psi \pi, \ \  \eta_c \rho, \ \ \psi(2S) \pi, \ \ D \bar D^*, \ \ D^* \bar D^*
\label{eq_channels}
\ee
Hence, the wave function has five components, 
the potential $V(\bm r) $ takes the form of $5 \times 5$  matrix, 
while the kinetic term is diagonal in the five channels.  
We have solved the coupled equation (\ref{eq_Schro_eq}) with suitable boundary conditions, 
and obtained the phase-shifts for diagonal channels ($\alpha \alpha$).  
Then they are transformed to scattering amplitudes by 
\be 
S_{\alpha \alpha}= e^{2i \delta_{\alpha \alpha}}, \ \  f_{\alpha \alpha} = \frac{1}{2ik_\alpha} (S_{\alpha \alpha}-1)
\label{eq_Sandf}
\ee
where $k_\alpha$ is the relative momentum of the  channel $\alpha$.

%---------------------------------
\subsection{\label{Potential} Potentials from the quark exchanges}
%---------------------------------

\begin{figure*}[t]
  \begin{center}
\includegraphics[width=\textwidth,clip]{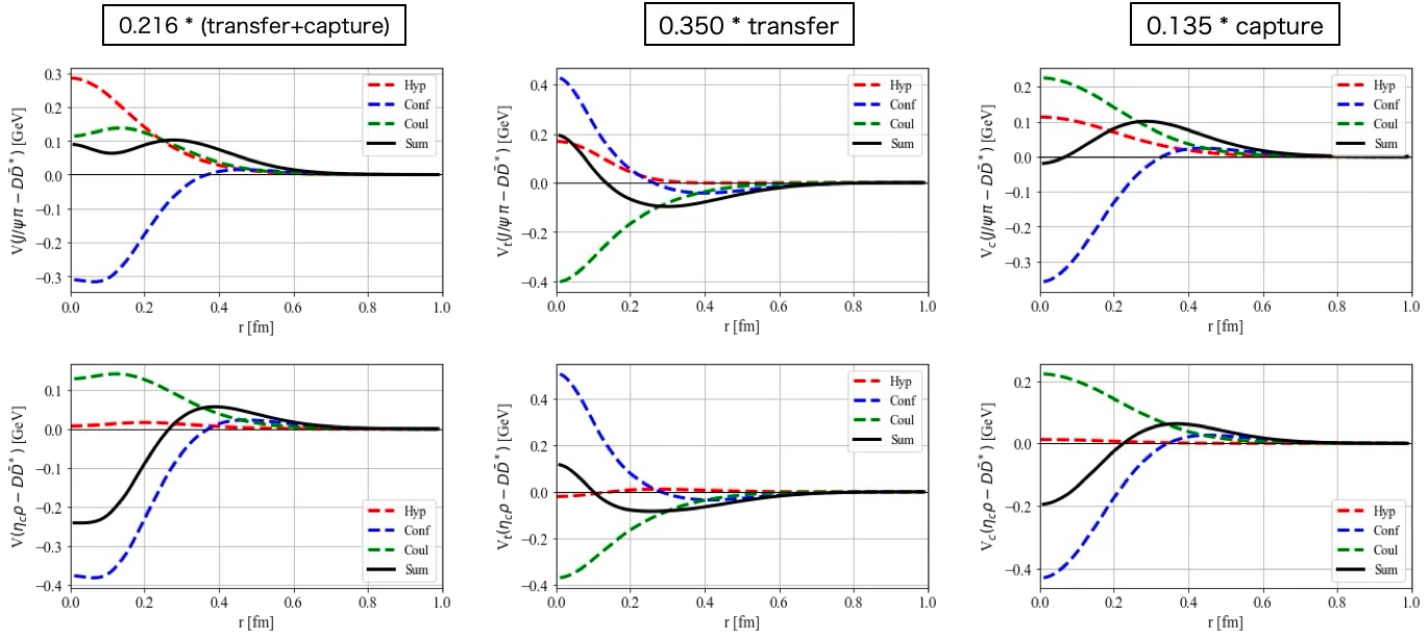}
\caption{
\label{fig_potential2} 
Important potentials for off-diagonal channels, $V(J/\psi\pi - D\bar D^*)$ (first row) and $V(\eta_c \rho - D\bar D^*)$ (second row) as functions of the radial distance $r$.  
Three different cases are shown depending on the terms included; the left column shows the results including transfer and capture processes,  the middle one transfer, and the right one capture.   These potentials are scaled by factors as shown in the top boxes in such a way that they reproduce the amplitudes the lattice calculation as shown in the next subsection.  }
  \end{center}  
 \end{figure*}

In this subsection we discuss the quark exchange potential in detail. 
As we will discuss in the next subsection, the potential strengths are 
reduced by about factor 0.3 to obtain better agreement with those 
of the lattice calculation.
As anticipated, there are various contributions from different diagrams as shown in 
Fig.~\ref{fig_q_exchange_full}.
To compare our results with lattice calculations, we show the two important transition potentials in Fig.~\ref{fig_potential2}, 
$V({J/\psi \pi {\text -} D\bar D^*})$ and 
$V({\eta_c \rho {\text -} D\bar D^*})$ as functions of 
radial distance $r$.
Three cases from left to right 
are Transfer + Capture (left), Transfer (middle) and Capture (right).  
Furthermore, contributions from the hyperfine (red dashed), confinement (blue dashed) and Coulomb terms 
(green dashed)
are shown separately.  
In all cases, the potential strengths are reduced by a factor as shown on the top of the figures.  
  
We observe the following points.  
\begin{itemize}

\item
Typical potential strength is about 0.2 GeV with short ranges; 
the potential almost vanishes at around 0.5 fm.  
For transition potentials, the sign of the potential is not important; only magnitude 
is relevant.  

\item
Although various components, hyperfine (Hyp), confinement (Conf) and Coulomb (Coul) terms, 
separately behave smoothly as functions of $r$, the sums show non-trivial $r$ dependences.  
This is because  different contributions have different strengths and ranges.  

\item
Although the $r$-dependence of the sum differs for different cases, 
volume integrals of the sum take similar values 
around 
\be
0.1 \sim 0.2 \ {\rm fm}^2 = 0.02 \sim 0.04 \ {\rm GeV} \cdot {\rm fm}^3
\ee
As we will see later, the volume integrals play an important role 
for the amplitudes of  low energy scattering.  
\end{itemize}

%-------------------------------
%-------------------------------

\begin{figure*}[t]
  \begin{center}
\includegraphics[width=1\textwidth,clip]{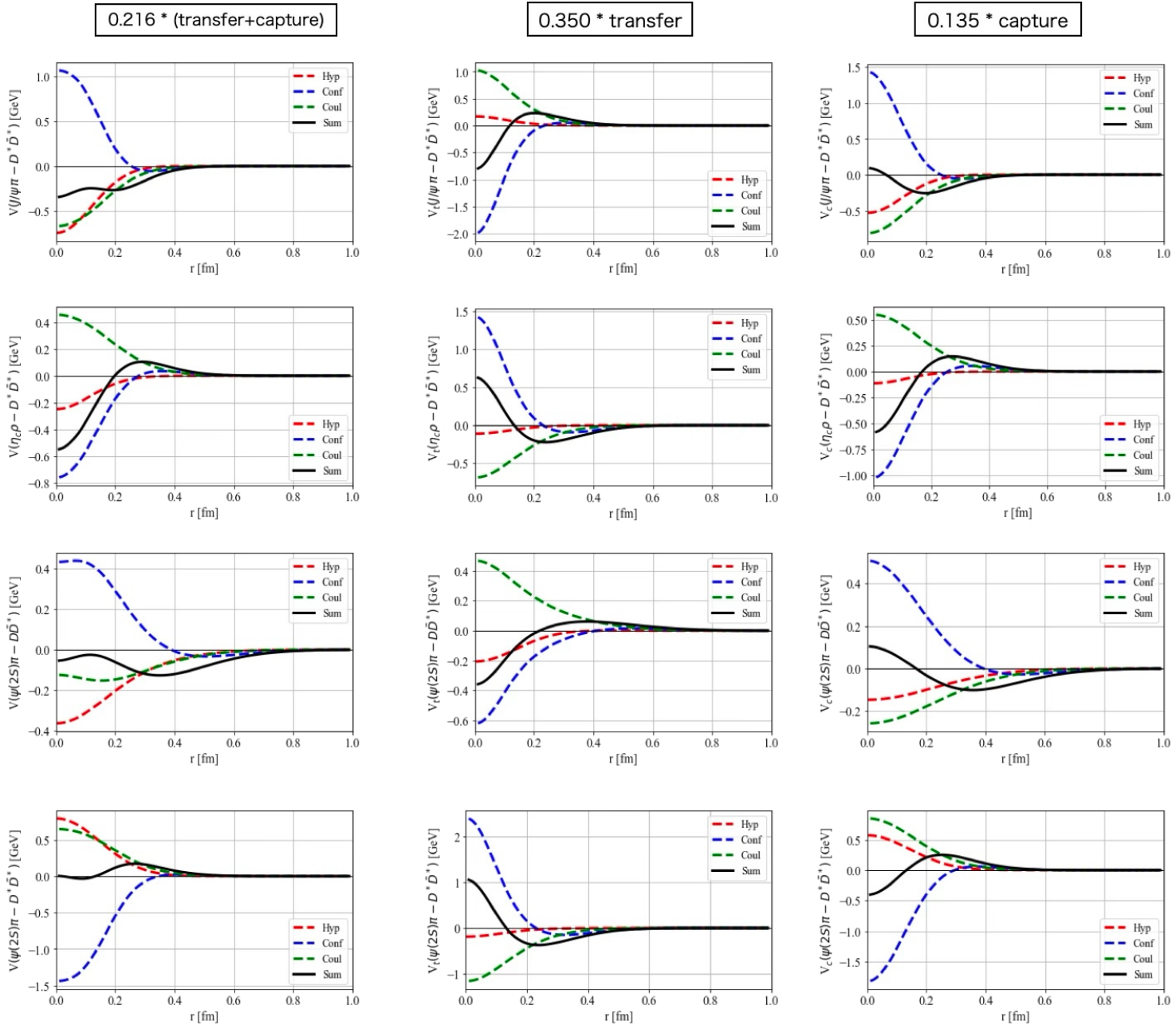}
\caption{
\label{fig_potential4} 
The other potentials generated by quark exchanges, $V(J/\psi\pi - D^*\bar D^*)$, $V(\eta_c \rho - D^*\bar D^*)$,  
$V(\psi(2S) \pi - D\bar D^*)$, $V(\psi(2S) \pi - D^*\bar D^*)$ from the top to bottom rows as functions of the radial distance $r$.  
The same classification and scales as in Fig.~\ref{fig_potential2} is used. }
  \end{center}  
 \end{figure*}

For quark exchange processes in our present model, 
there are four more potentials in addition to the above two, 
\be
& & V(J/\psi\pi\text{-}D^*\bar D^*),\ V(\eta_c \rho\text{-}D^*\bar D^*),
\label{eq_Vcomponents} \nonumber  \\
& & V(\psi(2S) \pi\text{-}D\bar D^*),\ V(\psi(2S) \pi\text{-}D^*\bar D^*) .
\ee
For completeness, we show these four potentials in Fig.~\ref{fig_potential4}.    
These potentials were not shown by the HALQCD in Refs.~\cite{Ikeda:2016zwx,Ikeda:2017mee}, because the $\psi(2S)\pi$ and $D^*\bar{D}^*$ channels were not explicitly included in their lattice calculations. However, as we will see later, these potentials provide non-negligible contributions in scattering amplitudes. 

%-------------------------------
%-------------------------------

%---------------------------------
\subsection{\label{Amplitudes} Amplitudes}
%---------------------------------

By using the obtained potential with the reduces strengths we have computed the scattering amplitudes of 
 the coupled channels
~(\ref{eq_channels}, \ref{eq_Sandf}), ${\rm Im} [f(E)]$, 
(the scattering amplitudes ${\rm Im} [f(E)]$ of the coupled channels in Eq.~\eqref{eq_channels})
where $E$ is the center-of-mass energy measured from the $J/\psi \pi$ threshold.  
To compare with the lattice results, we employ the masses of the particles, 
\be
& & m_{\pi} = 411\ {\rm MeV}, \ \ \ \  \ \ m_{\rho} = 896\ {\rm MeV}, 
\nonumber \\
& & m_{J/\psi} = 3097\ {\rm MeV}, \ \ m_{\eta_c} = 2988\ {\rm MeV}, 
\nonumber \\
& &  m_{D} = 1903\ {\rm MeV}, \ \ \ \  m_{D^*} = 2056\ {\rm MeV}
\label{eq_masses_HAL}
\ee
rather than the mass values in
Table~\ref{table_quarkmasses}.
By doing this threshold values of various channels agree with lattice results, which enables us to make better comparison between the two methods.

\begin{figure*}[t]
  \begin{center}
\includegraphics[width=0.95\textwidth,clip]{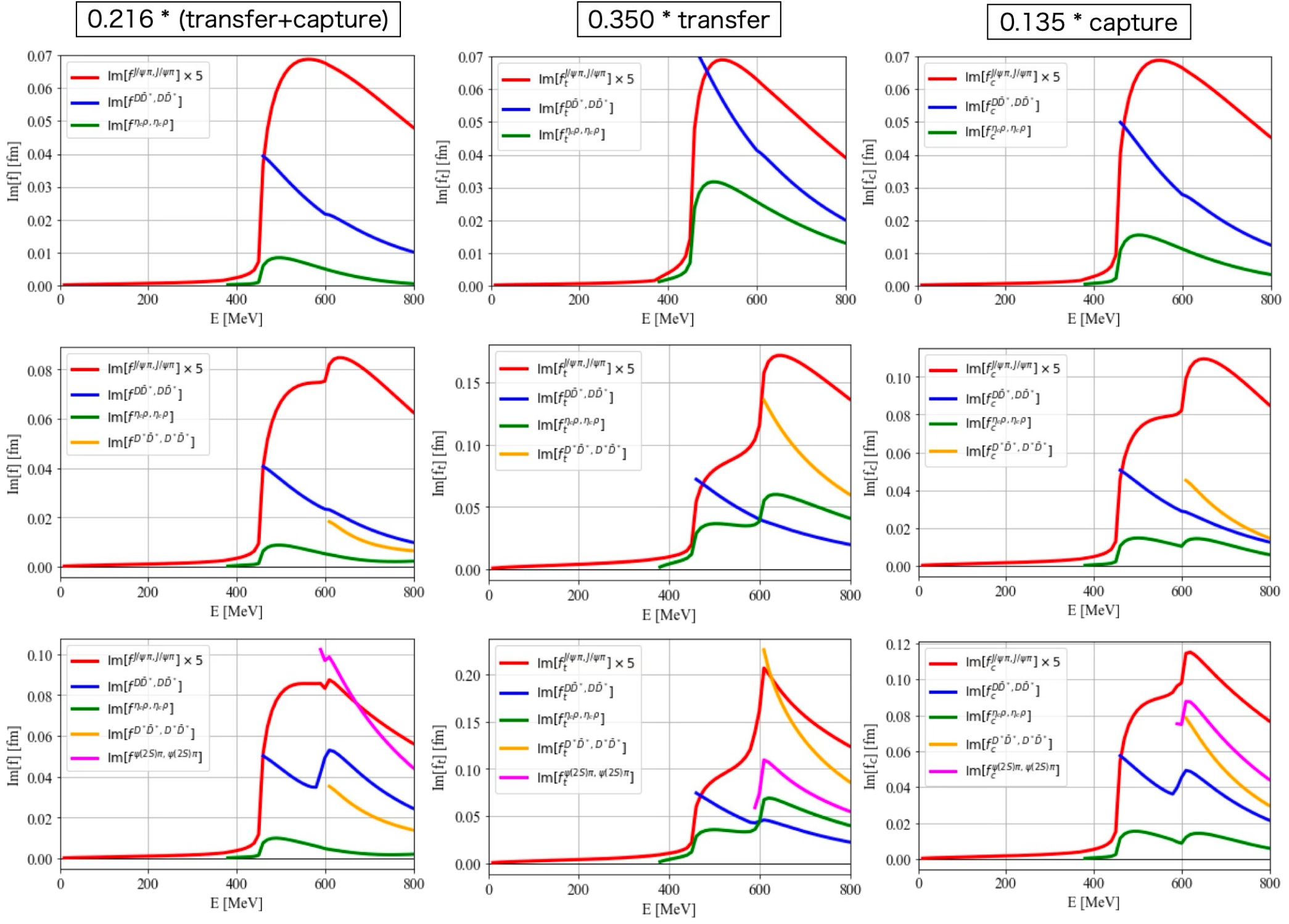}
\caption{
\label{fig_amplitudes} 
Various scattering amplitudes of the four channels, $J/\psi \pi$, $D\bar D^*$, $\eta_c \rho$ and $D^* \bar D^*$
as functions of the center-of-mass energy $E$ measured from the lowest threshold of  $J/\psi \pi$.
The arrangement of the three columns follows that of Fig.~\ref{fig_potential2}, while 
the three rows are for the amplitudes obtained by different coupled channels as shown in figures.
 }
  \end{center}  
 \end{figure*}

When we used the standard quark model parameters, the resulting amplitudes 
turned out to be too large which cannot be compared with the lattice result~\cite{Ikeda:2016zwx,Ikeda:2017mee}.
In fact, there are some ambiguities in the model setup and treatments.
One is the strength of the quark exchange potential  which is proportional to 
the parameter $\alpha_S$ .  
As in the standard quark model~\cite{Silvestre-Brac:1996myf}
the strength of $\alpha_S$ is determined in the light flavor sectors, 
where typical momentum scale is $\Lambda_{QCD}$.
Contrary, 
in the present quark exchange process it is larger of order 1 GeV.  
The coupling strength $\alpha_S$ scales to the leading order as
$1/\ln(Q^2/\Lambda_{QCD}^2) \equiv \beta$.
By setting $\Lambda_{QCD} \sim 0.2$ GeV, it is estimated as  
$\beta \sim 1/\ln 25 \sim 0.3$.  
We employ $\beta$ around this value as a uniform reduction factor,
\be
\tilde V(\bm r) = \beta V(\bm r)
\ee
In actual calculations we attempt several values around 0.3 
depending on the kind of quark exchange processes as indicated 
on top of Figs.~\ref{fig_potential2}, \ref{fig_potential4}, \ref{fig_amplitudes}.  
The other ambiguity is in the approximation in the potential structure.
In general,  the quark exchange amplitude becomes non-local.  
As explained in the previous section (Eqs. (31) and after), by neglecting the $\bm P$-dependence we have 
made the amplitude local.  
This prescription may effectively enhance the strength of the potential.  
To know better the non-local nature, we need a more sophisticated analysis 
which is a future work.

Results are shown in Fig.~\ref{fig_amplitudes} for 9 (= $3 \times 3$) cases.  
Three columns are for diagonal amplitudes of different exchange processes as corresponding to the potential plots in 
Fig.~\ref{fig_potential2},
and three rows are for three different cases of coupled channels; 
top panel: three coupled channels of $J/\psi\pi, D \bar D^*, \eta_c \rho$, 
middle panel: four of $J/\psi\pi, D \bar D^*, \eta_c \rho, D^* \bar D^* $, 
bottom panel: five of $J/\psi\pi, D \bar D^*, \eta_c \rho, D^* \bar D^*, \psi(2S)\pi$.

Let us compare the present result with that of the lattice calculations.  
As we can see in Fig.~\ref{fig_amplitudes} the top three plots are qualitatively similar to and consistent 
with the lattice result in the relative behaviors of the three amplitudes, see Figure 4 of Ref.~\cite{Ikeda:2017mee}.  
The reason for the similar behaviors of the three cases is that the volume integrals of the potentials take similar values as anticipated in the previous section.  
Consequently, as expected, our amplitude has no bound nor resonant poles in accordance with 
what has been claimed in Ref.~\cite{Ikeda:2017mee}.

Now as we increase the number of channels in our coupled channel treatment, 
line shapes of amplitudes exhibit kink-like singular behaviors at the threshold of the added channels, 
$D^* \bar D^*$ and $\psi(2S) \pi$, which is the requirement of the unitarity (probability conservation).  
Furthermore, absolute values of the amplitudes change significantly; as the number of channels increase
the amplitudes also increase.  
Therefore, proper treatment for coupled channel problems is very important.  
In comparison with the amplitudes of Ref.~\cite{Ikeda:2017mee}, they do not show singular like behaviors at threshold 
in the energy region higher than the $D \bar D^*$ threshold.
This discrepancy should be further considered in the future.

%-------------------------------
%-------------------------------

%---------------------------------
\subsection{\label{Toy} Toy model calculations}
%---------------------------------

To understand the behaviors of the amplitudes, we have performed a simple 
calculation using a three-channel model with a contact interaction.
The three important channels are for $J/\psi \pi$, $D \bar D^*$ and $\eta_c \rho$, 
which are labeled by $i = 1, 2, 3$, respectively.  
Hence the Hamiltonian is given by 
\be
H &=& 
\begin{pmatrix}
K_1 + V_{11} & V_{12} & V_{13} \\
V_{21} & K_2 + V_{22} & V_{23} \\
V_{31} & V_{32} & K_3+V_{33} 
\end{pmatrix}
\ee
where $K_i$ is the kinetic energy of the channel $i$, 
\be 
K_i = \frac{p_i^2}{2m_i}  
\to \frac{p_i^2}{2\mu_i} + \Delta M_i 
\ee
with the reduced mass $\mu_i$ and the threshold energy $\Delta M_i$ measured from the lowest threshold,
and the potential parameters $V_{ij} = V_{ji}$ are constant in momentum space, 
which is contact type in coordinate space.  
Because the contact interaction is separable, we can solve the 
Lippmann-Schwinger equation as an algebraic equation.
The resulting $t(E)$ matrix is given in momentum space by 
\be
t(E) = \frac{1}{1/V - G(E)}
\label{eq_t_sol}
\ee 
where the $G$-function (two-particle propagator) is given by 
\be
G(E) =
\int  \frac{d^3 q}{(2\pi)^3} \frac{f(\bm q)^2}{E - \bm q^2/2\mu}
\ee
with the reduced mass $\mu$ of the two particles.
To remove the ultraviolet divergence
we introduce a form factor $f(\bm q)$.  
Employing the monopole type for $f(\bm q)$ with the cutoff $\Lambda$, 
\be 
f(\bm q) = \frac{\Lambda^2}{\Lambda^2 + | \bm q|^2}
\ee
we obtain a simple expression for the $G$-function, 
\be
G(E) =
\frac{ \mu \Lambda^3 }{4 \pi} \frac{1}{(k + i \Lambda)^2}, \ \ \ k = \sqrt{2\mu E}
\label{eq_G0_compact}
\ee
The $t$-matrix here is related to the $f$-amplitude in 
(\ref{eq_Sandf}) by
\be
f(E) = - \frac{\mu}{2\pi} t(E)
\ee

Employing the mass values as used in the lattice calculation as in 
(\ref{eq_masses_HAL}), 
the parameters in this toy model are the potential strengths $V_{ij}$ and cutoff $\Lambda_i$.
We have attempted several parameter sets and a reasonable one is 
found to be as shown in Table~\ref{tbl_LVparameters}.
For the cutoff, we have employed different values for different channels $i$.
We note  that the value $\Lambda = 4\ {\rm fm} \sim 0.8 \ {\rm GeV}$
corresponds to a typical interaction range $R \sim \sqrt{6}/\Lambda \sim 0.6$ fm.  
For potential parameters, guided by the lattice result, we set only 
three components finite, $V_{12}, V_{22}, V_{23}$.
In fact, the interaction in the channel 1, $J/\psi \pi$, is expected to be suppressed.

%---------------------------------------
\begin{table}[htbp]
\caption{Parameters used in the toy model. 
The cutoff $\Lambda_i$'s are in units of fm, and the potentials $V_{ij}$ in units of fm$^2$.}
\begin{tabular}{cccccc}
\hline
$\Lambda_1$ & $\Lambda_2$ & $\Lambda_3$ & $V_{12}$ & $V_{22}$ & $V_{23}$ \\
4 & 3 & 3 & 0.23 & -0.12 & 0.21 \\ 
\hline
\end{tabular}
\label{tbl_LVparameters}
\end{table}%
%---------------------------------------

%-------------------------------
 \begin{figure}[t]
  \begin{center}
\includegraphics[width=0.8\linewidth,clip]{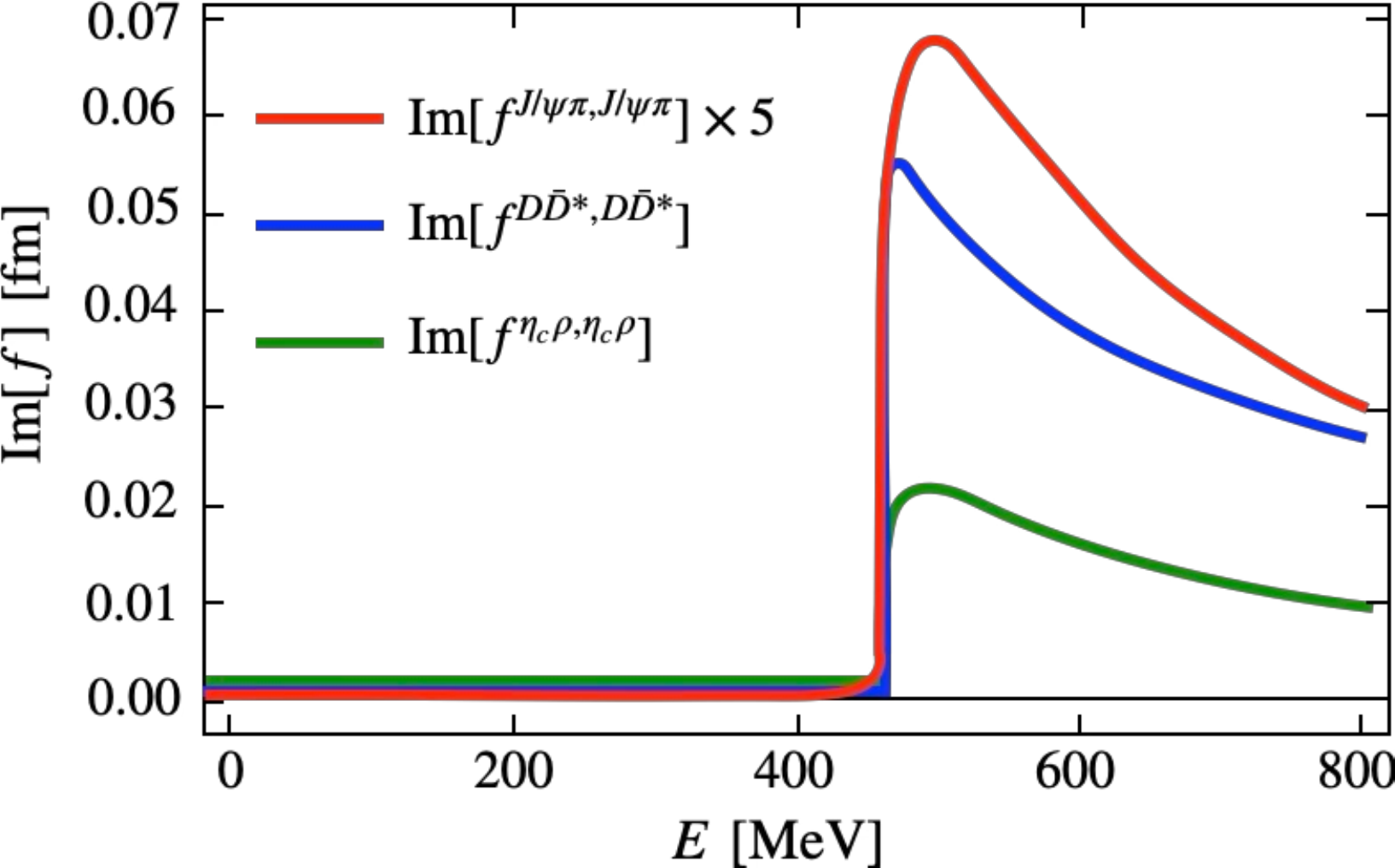}
\caption{
\label{fig_simple_f} 
   Various amplitude in a toy model.}
  \end{center}  
 \end{figure}
%-------------------------------

The result of the scattering amplitudes are shown in Fig.~\ref{fig_simple_f}.
We argue that the line shapes of the amplitudes are qualitatively consistent 
with those of the lattice~\cite{Ikeda:2016zwx,Ikeda:2017mee} 
and the present calculations.  
The potential strengths are also consistent with each other.  
The potential parameters $V_{ij}$ listed in Table~\ref{tbl_LVparameters} can be crudely identified with 
volume integrals of the (local) potential used in the lattice and present analysis.
This can be seen by a  momentum independent interaction which can be written as
\be
\bra \bm p^\prime | V | \bm p \ket = \bra \bm 0 | V | \bm 0 \ket 
=
\int d^3 x V(x) 
\ee
These volume integrals for the present quark model calculations are 
\be
V_{12} \sim 0.28, \ \ \ 
V_{23} \sim 0.17, \ \ \
V_{22} \sim 0
\ee
which are roughly consistent with those in Table~\ref{tbl_LVparameters}.  
We have computed the amplitudes with using other set of parameters.
It turned out that result depends very sensitively on the strengths of the $V_{ij}$ parameters
both in their relative ratios and absolute strengths.  
Observing this fact, the quark exchange mechanism in the quark model 
provides in a rather nontrivial way potential strengths 
that are consistent with the lattice calculation modulo overall strength.  
As we have discussed, the reduction of the overall strength could be attributed to the running coupling constant $\alpha_S$.  
Therefore, 
we could say that, though not perfect, the quark model provides 
a reasonable explanation for the phenomena around the $Z_c(3900)$ and useful to
understand physics mechanism of the lattice calculations.  
This observation also suggests  the suppression of the coupling of two-meson channels to 
genuine tetraquark state, and hence their existence as a real particle is not likely.  

%
%======================%
\section{\label{Sec:Summary} Summary and discussions}
%======================%
We have investigated hadron interactions between mesons in the $Z_c$ channels by meson exchange and quark exchange models.
While the quark exchange model has been employed for hadron-hadron interactions at short distances for
various systems, the case with constituents with different masses has been carefully formulated in the present work.  
We have seen that the meson exchanges play a negligible role, while the quark exchanges
at short distances dominate the interaction among various channels.  
In general the quark exchange processes yield a non-local interaction, which has been in the present work approximated by a local potential by neglecting higher order terms.  
The obtained potential is used in a coupled channel equation for the $Z_c$-channel including 
$J/\psi \pi$, $D \bar D^*$, $\eta_c \rho$, $D^* \bar D^*$ and $\psi(2S) \pi$.  
The resulting amplitudes are then compared with the results of the lattice simulations~\cite{Ikeda:2017mee}.  
We have shown that the amplitudes depend very sensitively on the interaction
by using a simple toy scattering model.  
It turns out that the quark exchange model explains reasonably well the results 
of the lattice simulations, 
once the strength of the obtained local potential is weakened uniformly, the amount of which 
is consistent with the running behavior of the strong coupling constant $\alpha_S$ between quarks and gluons.  

There are still several unsolved questions that have to be clarified.  
In comparison with the data, the use of the physical pion mass is important.
For this, we need updated lattice simulations and model calculations.  

Since the pioneering work of the $Z_c$ tetraquark in the lattice~\cite{Ikeda:2016zwx,Ikeda:2017mee}, 
there have been a great amount of progresses in hadron interactions.  
In the lattice simulations, the interactions in the $\phi N$, $J/\psi N$ channels have been 
achieved, revealing rather strong attractive interaction~\cite{Lyu:2025jjl,Lyu:2025jjl}.
This observation seems consistent with the experimental analysis 
of the correlation functions~\cite{ALICE:2021cpv}, 
while not with the photo-production analysis~\cite{Strakovsky:2020uqs}.
Theoretically, the importance of the two-pion exchange 
is recently revisited~\cite{Lyu:2025ohv,Fujii:1999xn,TarrusCastella:2018php}, 
which could be also relevant in the present study of the $Z_c$ channel.  
In the lattice calculations and in experimental data, the separation of spin states is also important
but not yet analyzed.  
In effective models better coupled channel models are expected including possible hadron-hadron channels, 
and also with genuine quark-dominated channels.  
These multi-face studies should be further explored in the elucidation of exotic hadron phenomena.

%================================
\section*{Acknowledgments}
%================================

We thank Kenji Fukukawa and Yoichi Ikeda for discussions and supports.
This work is supported in part by Grants-in Aid for Scientific Research on Innovative Areas, No. 18H05407, and No. 24K07050(C) (A.H.), and JP20K14478 (Y.Y.). 
The numerical calculations were carried out on heian at YITP in Kyoto University.

\appendix

%======================%
\section{\label{appendix:Vectorexchange} Vector $\rho$ and $\omega$ meson exchange}
%======================%

The necessary $VD^{(*)}D^{(*)}$ vertices are extracted from (\ref{L_VHH}).  
They are
\be
\calL_{VDD} &=& - \beta g_V D v_\mu V^\mu D^\dagger
\label{L_VDD2} \\
\calL_{VD^*D^*} &=& \beta g_V D^{*\mu} v_\mu V^\mu D^{* \mu \dagger}
\nonumber \\
&+&  2i \lambda g_V D^{*\mu}  F_{\mu \nu} D^{*\nu \dagger}
\label{L_VD*D*2} \\
\calL_{VDD^*} &=& -i 2 \lambda g_V 
\epsilon^{\alpha \beta \mu \nu} v_\alpha 
\nonumber \\
&\times& \left( D \del_\mu V_\nu D^{* \dagger}_\beta
- D^{*}_\beta \del_\mu V_\nu D^\dagger \right)
\label{L_VDD*2}
\ee
where $D$ meson isospin doublets are, 
\be
D = (D^0, D^+)
\ee
The resulting potentials for the $\rho$ meson exchange are
\begin{align}
 V^\rho_{D\bar{D}-D\bar{D}}(r)=&\left(\frac{\beta g_V}{2m_\rho}\right)^2C(r;m_\rho)\bm{\tau}_1\cdot\bm{\tau}_2,\\
 V^\rho_{D\bar{D}^\ast-D\bar{D}^\ast}(r)=&\left(\frac{\beta g_V}{2m_\rho}\right)^2\bm{\varepsilon}\,^\ast\cdot\bm{\varepsilon}\,C(r;m_\rho)\bm{\tau}_1\cdot\bm{\tau}_2,
\end{align}
\be
& & V^\rho_{D\bar{D}^\ast-D^\ast\bar{D}}(r) =(\lambda g_V)^2\frac{1}{3}
 [ -2\bm{\varepsilon}\,^\ast\cdot\bm{\varepsilon}\,D(r) 
 \nonumber \\
& &\ \ \ + 2\bm{\varepsilon}\,^\ast\cdot\bm{\varepsilon}\,C(r;\mu_\rho)
 -S_{\varepsilon\varepsilon}(\hat{r})T(r;\mu_\rho)
] \bm{\tau}_1\cdot\bm{\tau}_2, \\
& &  V^\rho_{D\bar{D}-D^\ast\bar{D}^\ast}(r)=-(\lambda g_V)^2
\frac{1}{3}[ 
 -2\bm{\varepsilon}\,^\ast\cdot\bm{\varepsilon}\,^\ast D(r)
 \nonumber \\
 & & \ \ \ +2\bm{\varepsilon}\,^\ast\cdot\bm{\varepsilon}\,^\ast C(r;m_\rho)
 -S_{\varepsilon\varepsilon}(\hat{r})T(r;m_\rho)
] \bm{\tau}_1\cdot\bm{\tau}_2, \\
& &  V^\rho_{D\bar{D}^\ast-D^\ast\bar{D}^\ast}(r)=i(\lambda g_V)^2\frac{1}{3}
[
 -2\bm{\varepsilon}\,^\ast\cdot\bm{S}\, D(r)
 \nonumber \\
& & \ \ \  +2\bm{\varepsilon}\,^\ast\cdot\bm{S}\, C(r;m_\rho)
 -S_{\varepsilon S}(\hat{r})T(r;m_\rho)
]\bm{\tau}_1\cdot\bm{\tau}_2, \\
& & V^\rho_{D^\ast\bar{D}^\ast-D^\ast\bar{D}^\ast}(r)=
 \left(\frac{\beta g_V}{2m_\rho}\right)^2\bm{\varepsilon}\,^\ast\cdot\bm{\varepsilon}\,
 \bm{\varepsilon}\,^\ast\cdot\bm{\varepsilon}\, C(r;m_\rho)\bm{\tau}_1\cdot\bm{\tau}_2 \notag\\
& &\ \ \ +(\lambda g_V)^2\frac{1}{3}\left[
 -2\bm{S}\,\cdot\bm{S}\, D(r) \right. 
 \nonumber \\
 & & \ \ \ \left.
 +2\bm{S}\,\cdot\bm{S}\, C(r;m_\rho)
 -S_{S S}(\hat{r})T(r;m_\rho)
 \right]\bm{\tau}_1\cdot\bm{\tau}_2,
\ee
For the $\omega$ meson exchange, $\bm{\tau}_1\cdot\bm{\tau}_2$ is removed.  

%======================%
\section{\label{appendix:Dexchange} $D^{(*)}$ meson exchange}
%======================%

The relevant vertex Lagrangians involving 
$J/\psi$ or $\psi^\prime$ in Fig.~\ref{fig_meson_exchange} are
\begin{align}
 {\cal L}_{\psi D\bar{D}}& = 
 2g^\prime \psi^\mu \left[\bar{D}^\dagger (\partial_\mu D^\dagger)
 - (\partial_\mu \bar{D}^\dagger)  D^\dagger \right]
 \notag\\
 & \quad + 2g^\prime \left[
 (\partial_\mu D)  \bar{D}
 - D (\partial_\mu \bar{D})   
 \right]\psi^{\mu\dagger}
 \\[2mm]
 {\cal L}_{\psi D\bar{D}^\ast} 
 & = 
 2g^\prime \varepsilon^{ijk} \psi_i \left[
 \bar{D}^{\ast \dagger}_j (\partial_k D^\dagger)
 - (\partial_k\bar{D}^{\ast \dagger}_j)  D^\dagger
 \right]
 \notag\\
 &\quad 
 + 2g^\prime \varepsilon^{ijk}\left[
 (\partial_k D)  \bar{D}^{\ast}_j
 - D (\partial_k \bar{D}^{\ast}_j) 
 \right] \psi^{\dagger}_i
 \\[2mm]
 {\cal L}_{\psi D^\ast \bar{D}}& = 
 2g^\prime \varepsilon^{ijk} \psi_i \left[
 \bar{D}^\dagger (\partial_j D^{\ast\dagger}_{k})
 - (\partial_j \bar{D}^\dagger) D^{\ast\dagger}_{k}
 \right]
 \notag\\
 &\quad + 2g^\prime \varepsilon^{ijk}\left[
 (\partial_j D^\ast_k)  \bar{D}
 - D^\ast_k (\partial_j \bar{D}) 
 \right] \psi^\dagger_i
 \end{align}
 \begin{align}
 {\cal L}_{\psi D^\ast \bar{D}^\ast}& = 
 2g^\prime \psi^\mu [
 \bar{D}^{\ast\dagger}_\mu (\partial^\nu D^{\ast \dagger}_\nu)
 - (\partial^\nu \bar{D}^{\ast\dagger}_\mu)  D^{\ast \dagger}_\nu
\nonumber \\
& - ( 
 \bar{D}^{\ast \nu \dagger} (\partial_\mu D^{\ast \dagger}_\nu)
 - (\partial_\mu \bar{D}^{\ast\nu\dagger})  D^{\ast \dagger}_\nu
 )
 \nonumber \\
& + \bar{D}^{\ast\dagger}_\nu (\partial^\nu D^{\ast \dagger}_\mu)
 - (\partial^\nu \bar{D}^{\ast\dagger}_\nu)  D^{\ast \dagger}_\mu
]
 \notag\\
 & + 2g^\prime
[
 (\partial^\nu D^{\ast}_\nu)\bar{D}^{\ast}_\mu
 - D^{\ast}_\nu (\partial^\nu \bar{D}^{\ast}_\mu)
 \nonumber \\
 & -( (\partial_\mu D^\ast_\nu)\bar{D}^{\ast \nu} 
 - D^\ast_\nu(\partial_\mu\bar{D}^{\ast\nu})  )
 \nonumber \\
& +(\partial^\nu D^{\ast}_\mu)\bar{D}^{\ast}_\nu 
 - D^{\ast}_\mu (\partial^\nu \bar{D}^{\ast}_\nu )
]\psi^{\mu\dagger}
\end{align}
and for those involving $\eta_c$ are 
 \begin{align}
 {\cal L}_{\eta_c D\bar{D}^\ast} & = 
 -2g^\prime \eta_c \left[
 \bar{D}^{\ast\mu\dagger} (\partial_\mu D^\dagger) 
 - (\partial_\mu \bar{D}^{\ast\mu\dagger})  D^\dagger
 \right]
 \notag\\
 & \quad -2g^\prime \left[
 (\partial_\mu D)\bar{D}^{\ast \mu} 
 -  D (\partial_\mu \bar{D}^{\ast \mu} )
 \right] \eta_c^\dagger
 \\[2mm]
 {\cal L}_{\eta_c D^\ast \bar{D}} & = 
 - 2g^\prime \eta_c \left[
 \bar{D}^{\dagger} (\partial_\mu D^{\ast \mu \dagger}) 
 -(\partial_\mu \bar{D}^{\dagger}) D^{\ast \mu \dagger}
 \right]
 \notag\\
 & \quad 
 - 2 g^\prime \left[ (\partial_\mu D^{\ast \mu})\bar{D} 
 - D^{\ast\mu} (\partial_\mu \bar{D})
 \right] \eta_c^\dagger
 \\[2mm]
 {\cal L}_{\eta_c D^\ast \bar{D}^\ast} & = 
 -2g^\prime v_0 \eta_c \varepsilon^{ijk} \left[
 \bar{D}^{\ast \dagger}_i (\partial_j D^{\ast \dagger}_k)
 - (\partial_j \bar{D}^{\ast \dagger}_i) D^{\ast \dagger}_k
 \right]
 \notag\\
 & \quad 
 -2g^\prime v_0 \left[
 (\partial_j D^\ast_k)\bar{D}^\ast_i 
 - P^\ast_k (\partial_j \bar{D}^\ast_i )
 \right] \eta_c^\dagger
\end{align}

The exchanging of a $D^{(*)}$ meson is accompanied by a large energy transfer.  
Therefore, we follow a somewhat primitive method of computing 
the transition amplitude.  
To be specific, let us consider the transition amplitude shown 
in Fig.~\ref{fig_meson_exchange} by the 
second order perturbation method, 
\be
\tilde V_{D \bar D^* \to J/\psi \pi}^D
&=& \sum_n \frac{\langle f | \calL_{\pi HH} | n \rangle \langle n | \calL_{\psi HH} | i \rangle}{E_i - E_n}
\nonumber \\
&=&
- \frac{1}{2m_D(2m_{J/\psi}2E_\pi(\bm p^\prime))^{1/2}} 
\nonumber \\
&\times&
\left( \frac{g_\pi}{f_\pi}m_D \right) 
\left( 2 g^\prime m_D m_{J/\psi}^{1/2} \right) 
\nonumber \\
&\times&
\frac{1}{2m_D}
\frac{\langle f | \bm S_1 \cdot \bm q | n \rangle \langle n | \bm S_2 \cdot \bm q | i \rangle}{E_\pi(\bm p^\prime)}, 
\ee
where $E_i, E_n$ are the energies of the initial and intermediate states.  
In the second step we have shown explicitly various factors; 
the first one $1/2m_D(2m_{J/\psi}2E_\pi(\bm p^\prime))^{1/2}$ is for the normalization per unit volume, 
the second and third one are the coupling constants with the heavy hadron field 
normalization taken into account, 
and the last one of $1/2m_D$ is for the normalization of the exchanged $D$ meson.  
The intermediate state $n$ is for the virtual $D$ exchange, 
and 
$\bm S_1, \bm S_2$ are suitable spin transition operators at the vertices 
with $\bm q = \bm p^\prime - \bm p$.  
In addition to this diagram, we need to consider $D^+D^{*-}$ component to form the isospin 1 
and charge conjugation eigenstate as in (\ref{CCeigenstate}).  
We find  
\be
\tilde V_{D \bar D^* \to J/\psi \pi}
=
- \frac{g_\pi g^\prime }{\sqrt{2}f_\pi E_\pi^{3/2}} 
 \bm S_1 \cdot \bm q  \bm S_2 \cdot \bm q 
\label{VDD*psipi}
\ee
where $E_\pi = 2m_D - m_{J/\psi}$ as we have discussed around Eq.~(\ref{Epi}).  
Similarly, we find the $D^*$ exchange potential.
This result is due to heavy quark symmetry.  

%======================%
\section{\label{appendix:Spinfactor} Spin factor}
%======================%
The spin factors $I_{spin}^{(i)}$ ($i=c1,c2,c3,c4,t1,t2$) in Eq.~\eqref{eq:BornAmp_1} are summarized in Tables~\ref{table:spinfactor_I}-\ref{table:spinfactor_t2}.

\begin{table}[htbp]
 \caption{\label{table:spinfactor_I} Spin factors of the spin-independent potentials for the $(S_A,S_B) \to (S_C,S_D)$ transition, where $S_i$ ($i=A,B,C,D$) is the spin of the meson $i$.}
 \begin{center}
  \begin{tabular}{c|ccc}
   \toprule[0.3mm]
    &$(1_A,1_B)$&$(1_A,0_B)$&$(0_A,1_B)$ \\ \hline
   $(1_C,1_D)$& $0$ & $-\frac{1}{\sqrt{2}}$ & $\frac{1}{\sqrt{2}}$ \\ 
 $(1_C,0_D)$&$-\frac{1}{\sqrt{2}}$ & $\frac{1}{2}$ & $\frac{1}{2}$ \\ 
 $(0_C,1_D)$&$\frac{1}{\sqrt{2}}$ & $\frac{1}{2}$ & $\frac{1}{2}$ \\
   \bottomrule[0.3mm]
  \end{tabular}
 \end{center}
\end{table}

\begin{table}[htbp]
 \caption{\label{table:spinfactor_c1}
 Spin factors of the spin-dependent potentials for the capture 1.}
 \begin{center}
  \begin{tabular}{c|ccc}
   \toprule[0.3mm]
    &$(1_A,1_B)$&$(1_A,0_B)$&$(0_A,1_B)$ \\ \hline
   $(1_C,1_D)$ & $0$ & $-\frac{1}{4 \sqrt{2}}$ & $\frac{1}{4 \sqrt{2}}$ \\
 $(1_C,0_D)$& $-\frac{1}{4 \sqrt{2}}$ & $\frac{1}{8}$ & $\frac{1}{8}$ \\
$(0_C,1_D)$&  $-\frac{3}{4 \sqrt{2}}$ & $-\frac{3}{8}$ & $-\frac{3}{8}$ \\
   \bottomrule[0.3mm]
  \end{tabular}
 \end{center}
\end{table}

\begin{table}[htbp]
 \caption{\label{table:spinfactor_c2} 
  Spin factors of the spin-dependent potentials for the capture 2.}
 \begin{center}
  \begin{tabular}{c|ccc}
   \toprule[0.3mm]
    &$(1_A,1_B)$&$(1_A,0_B)$&$(0_A,1_B)$ \\ \hline
   $(1_C,1_D)$ & $0$ & $-\frac{1}{4 \sqrt{2}}$ & $\frac{1}{4 \sqrt{2}}$ \\
   $(1_C,0_D)$& $\frac{3}{4 \sqrt{2}}$ & $-\frac{3}{8}$ & $-\frac{3}{8}$ \\
   $(0_C,1_D)$& $\frac{1}{4 \sqrt{2}}$ & $\frac{1}{8}$ & $\frac{1}{8}$ \\
   \bottomrule[0.3mm]
  \end{tabular}
 \end{center}
\end{table}

\begin{table}[htbp]
 \caption{\label{table:spinfactor_c3}
   Spin factors of the spin-dependent potentials for the capture 3.}
 \begin{center}
  \begin{tabular}{c|ccc}
   \toprule[0.3mm]
    &$(1_A,1_B)$&$(1_A,0_B)$&$(0_A,1_B)$ \\ \hline
   $(1_C,1_D)$ &
	       $0$ & $-\frac{1}{4\sqrt{2}}$ &  $-\frac{3}{4\sqrt{2}}$ 
	       \\
 $(1_C,0_D)$& 
	       $-\frac{1}{4\sqrt{2}}$ & $\frac{1}{8}$ & $-\frac{3}{8}$
	       \\
$(0_C,1_D)$&  
	       $\frac{1}{4\sqrt{2}}$ 
	       & $\frac{1}{8}$ & $-\frac{3}{8}$
	       \\
   \bottomrule[0.3mm]
  \end{tabular}
 \end{center}
\end{table}

\begin{table}[htbp]
 \caption{\label{table:spinfactor_c4} 
  Spin factors of the spin-dependent potentials for the capture 4. }
 \begin{center}
  \begin{tabular}{c|ccc}
   \toprule[0.3mm]
    &$(1_A,1_B)$&$(1_A,0_B)$&$(0_A,1_B)$ \\ \hline
   $(1_C,1_D)$ &
       $0$ & $\frac{3}{4\sqrt{2}}$ & $\frac{1}{4\sqrt{2}}$
	       \\
 $(1_C,0_D)$&
       $-\frac{1}{4\sqrt{2}}$ & $-\frac{3}{8}$ & $\frac{1}{8}$
	       \\
$(0_C,1_D)$&  
       $\frac{1}{4\sqrt{2}}$ & $-\frac{3}{8}$ & $\frac{1}{8}$
	       \\
   \bottomrule[0.3mm]
  \end{tabular}
 \end{center}
\end{table}

\begin{table}[htbp]
 \caption{\label{table:spinfactor_t1}   Spin factors of the spin-dependent potentials for the transfer 1.}
 \begin{center}
  \begin{tabular}{c|ccc}
   \toprule[0.3mm]
    &$(1_A,1_B)$&$(1_A,0_B)$&$(0_A,1_B)$ \\ \hline
   $(1_C,1_D)$ &$-\frac{1}{2}$ & $\frac{1}{4 \sqrt{2}}$ & $-\frac{1}{4 \sqrt{2}}$ \\
   $(1_C,0_D)$& $\frac{1}{4 \sqrt{2}}$ & $-\frac{1}{8}$ & $\frac{3}{8}$ \\
   $(0_C,1_D)$&$-\frac{1}{4 \sqrt{2}}$ & $\frac{3}{8}$ & $-\frac{1}{8}$ \\
   \bottomrule[0.3mm]
  \end{tabular}
 \end{center}
\end{table}

\begin{table}[htbp]
 \caption{\label{table:spinfactor_t2} Spin factors of the spin-dependent potentials for the transfer 2.}
 \begin{center}
  \begin{tabular}{c|ccc}
   \toprule[0.3mm]
    &$(1_A,1_B)$&$(1_A,0_B)$&$(0_A,1_B)$ \\ \hline
   $(1_C,1_D)$ &  $\frac{1}{2}$ & $\frac{1}{4 \sqrt{2}}$ & $-\frac{1}{4 \sqrt{2}}$ \\
   $(1_C,0_D)$& $\frac{1}{4 \sqrt{2}}$ & $\frac{3}{8}$ & $-\frac{1}{8}$ \\
   $(0_C,1_D)$& $-\frac{1}{4 \sqrt{2}}$ & $-\frac{1}{8}$ & $\frac{3}{8}$ \\
   \bottomrule[0.3mm]
  \end{tabular}
 \end{center}
\end{table}

\section{\label{appendix:BornAmp} Summary of the 
Born
amplitudes}
In this appendix, the Born amplitude $\bra C,D|v_{ij}|A,B \ket$ of each of the diagrams is summarized.

For the capture 1, 
\begin{align}
 & \bra {C,D|v_{a\bar{b}}|A,B} \ket \notag\\
    &=  \,
 \left(-\frac{4}{9}\right) 
 I_{spin}^{(c1)}
 \frac{-1}{(2\pi)^3} 
 \delta^{(3)}(\bm{C}+\bm{D}-\bm{A}-\bm{B})
 \notag\\
 & \quad \times 
 \int d^3a \,d^3c  
 \, 
 \Phi_{C}^\ast(\bm{c_r}^{(c1)})\Phi_{D}^\ast(\bm{d_r}^{(c1)})
 v_{a\bar{b}}(|\bm{c}-\bm{a}|) \notag\\
 & \quad \times \Phi_{A}(\bm{a_r}^{(c1)})\Phi_{B}(\bm{b_r}^{(c1)}) 
 \notag\\
 & \equiv (2\pi)^3 \delta^{(3)}(\bm{C}+\bm{D}-\bm{A}-\bm{B}) h^{(c1)}_{fi} ,
\end{align}
where the meson momenta are given by
\begin{align}
 \bm{a_r}^{(c1)} &
 = \bm{a} - \frac{m_a}{m_a+\bar{m}_a}\bm{A} ,
 \\
 %%%
 \bm{b_r}^{(c1)} &
 = \bm{a} - \bm{C} - \frac{\bar{m}_b}{m_b+\bar{m}_b}\bm{A} ,
 \\
 %%%
  \bm{c_r}^{(c1)} &
 = \bm{c} - \frac{m_c}{m_c+\bar{m}_c}\bm{C} ,
 \\
 %%%
 \bm{d_r}^{(c1)} &
 = \bm{a} - \bm{A} - \frac{\bar{m}_d}{m_d+\bar{m}_d}\bm{C} . 
\end{align}

For the capture 2, 
\begin{align}
& \bra {C,D|v_{\bar{a}b}|A,B} \ket \notag\\
    &=  \,
 \left(-\frac{4}{9}\right) 
 I_{spin}^{(c2)}
 \frac{-1}{(2\pi)^3} 
 \delta^{(3)}(\bm{C}+\bm{D}-\bm{A}-\bm{B})
 \notag\\
 &\quad \times \int d^3\bar{a} \,d^3\bar{d}  
 \, 
 \Phi_{C}^\ast(\bm{c_r}^{(c2)})\Phi_{D}^\ast(\bm{d_r}^{(c2)})
 v_{\bar{a}b}(|\bm{\bar{d}}-\bm{\bar{a}}|) \notag\\
 & \quad \times \Phi_{A}(\bm{a_r}^{(c2)})\Phi_{B}(\bm{b_r}^{(c2)}) ,
 \notag\\
 & \equiv (2\pi)^3 \delta^{(3)}(\bm{C}+\bm{D}-\bm{A}-\bm{B}) h^{(c2)}_{fi} ,
\end{align}
where 
\begin{align}
 \bm{a_r}^{(c2)} &
 = - \bm{\bar{a}} + \frac{\bar{m}_a}{m_a+\bar{m}_a}\bm{A} ,
 \\
 %%%
 \bm{b_r}^{(c2)} &
 = -\bm{\bar{a}} - \bm{C} + \frac{m_b}{m_b+\bar{m}_b}\bm{A} ,
 \\
 %%%
  \bm{c_r}^{(c2)} &
 = -\bm{\bar{a}} + \bm{A} - \frac{m_c}{m_c+\bar{m}_c}\bm{C} ,
 \\
 %%%
 \bm{d_r}^{(c2)} &
 = -\bm{\bar{d}} - \frac{\bar{m}_d}{m_d+\bar{m}_d}\bm{C} .
\end{align}

For the capture 3, 
\begin{align}
 & \bra{C,D|v_{c\bar{d}}|A,B}\ket \notag\\
    &=  \,
 \left(-\frac{4}{9}\right) 
 I_{spin}^{(c3)}
 \frac{-1}{(2\pi)^3} 
 \delta^{(3)}(\bm{C}+\bm{D}-\bm{A}-\bm{B})
 \notag\\
 & \times 
 \int d^3{a} \,d^3{c}  
 \, 
 \Phi_{C}^\ast(\bm{c_r}^{(c3)})\Phi_{D}^\ast(\bm{d_r}^{(c3)})
 v_{c\bar{d}}(|\bm{c}-\bm{a}|) \notag\\
 & \quad \times \Phi_{A}(\bm{a_r}^{(c3)})\Phi_{B}(\bm{b_r}^{(c3)}) 
 \notag\\
 & \equiv (2\pi)^3
 \delta^{(3)}(\bm{C}+\bm{D}-\bm{A}-\bm{B}) h_{fi}^{(c3)} ,
\end{align}
where 
\begin{align}
 \bm{a_r}^{(c3)} &
 =  \bm{a} - \frac{m_a}{m_a+\bar{m}_a}\bm{A} ,
 \\
 %%%
 \bm{b_r}^{(c3)} &
 = \bm{c} - \bm{C} - \frac{\bar{m}_b}{m_b+\bar{m}_b}\bm{A} ,
 \\
 %%%
  \bm{c_r}^{(c3)} &
 = \bm{c} - \frac{m_c}{m_c+\bar{m}_c}\bm{C} ,
 \\
 %%%
 \bm{d_r}^{(c3)} &
 = \bm{c} - \bm{A} - \frac{\bar{m}_d}{m_d+\bar{m}_d}\bm{C} .
\end{align}

For the capture 4, 
\begin{align}
 &\bra {C,D|v_{\bar{c}d}|A,B} \ket \notag\\
 &=  \,
 \left(-\frac{4}{9}\right) 
 I_{spin}^{(c4)}
 \frac{-1}{(2\pi)^3} 
 \delta^{(3)}(\bm{C}+\bm{D}-\bm{A}-\bm{B})
 \notag\\
 & \quad \times 
 \int d^3{a} \,d^3{c}  
 \, 
 \Phi_{C}^\ast(\bm{c_r}^{(c4)})\Phi_{D}^\ast(\bm{d_r}^{(c4)})
 v_{\bar{c}d}(|\bm{c}-\bm{a}|) \notag\\
 & \quad \times\Phi_{A}(\bm{a_r}^{(c4)})\Phi_{B}(\bm{b_r}^{(c4)}) 
 \notag\\
 & \equiv
 (2\pi)^3
 \delta^{(3)}(\bm{C}+\bm{D}-\bm{A}-\bm{B}) h_{fi}^{(c4)} ,
\end{align}
where
\begin{align}
 \bm{a_r}^{(c4)} &
 =  \bm{d} +\bm{C} + \frac{\bar{m}_a}{m_a+\bar{m}_a}\bm{A} ,
 \\
 %%%
 \bm{b_r}^{(c4)} &
 = \bm{b} + \frac{m_b}{m_b+\bar{m}_b}\bm{A} ,
 \\
 %%%
  \bm{c_r}^{(c4)} &
 = \bm{d} +\bm{A} + \frac{\bar{m}_c}{m_c+\bar{m}_c}\bm{C} ,
 \\
 %%%
 \bm{d_r}^{(c4)} &
 = \bm{d} + \frac{m_d}{m_d+\bar{m}_d}\bm{C} .
\end{align}

For the transfer 1, 
\begin{align}
 & \bra {C,D|v_{ab}|A,B} \ket \notag\\ &=  \,  
 \left(+\frac{4}{9}\right) 
 I_{spin}^{(t1)}
 \frac{-1}{(2\pi)^3} 
 \delta^{(3)}(\bm{C}+\bm{D}-\bm{A}-\bm{B})
 \notag\\
 &\quad \times \int d^3a \,d^3c  
 \, 
 \Phi_{C}^\ast(\bm{c_r}^{(t1)})\Phi_{D}^\ast(\bm{d_r}^{(t1)})
 v_{ab}(|\bm{c}-\bm{a}|) \notag\\
 & \quad \times \Phi_{A}(\bm{a_r}^{(t1)})\Phi_{B}(\bm{b_r}^{(t1)}) ,
 \notag\\
 & \equiv (2\pi)^3 \delta^{(3)}(\bm{C}+\bm{D}-\bm{A}-\bm{B}) h^{(t1)}_{fi} ,
\end{align}
where 
\begin{align}
 \bm{a_r}^{(t1)} &
 = \bm{a} - \frac{m_a}{m_a+\bar{m}_a}\bm{A} ,
 \\
 %%%
 \bm{b_r}^{(t1)} &
 = \bm{c} - \bm{C} - \frac{\bar{m}_b}{m_b+\bar{m}_b}\bm{A} ,
 \\
 %%%
  \bm{c_r}^{(t1)} &
 = \bm{c} - \frac{m_c}{m_c+\bar{m}_c}\bm{C} ,
 \\
 %%%
 \bm{d_r}^{(t1)} &
 = \bm{a} - \bm{A} - \frac{\bar{m}_d}{m_d+\bar{m}_d}\bm{C} .
\end{align}

For the transfer 2,
\begin{align}
 & \bra {C,D|v_{\bar{a}\bar{b}}|A,B} \ket \notag\\
  & = \, 
 \left(+\frac{4}{9}\right) 
 I_{spin}^{(t2)}   
 \frac{-1}{(2\pi)^3} 
 \delta^{(3)}(\bm{C}+\bm{D}-\bm{A}-\bm{B})
 \notag\\
 & \quad \times \int d^3\bar{a} \,d^3\bar{d}  
 \, 
 \Phi_{C}^\ast(\bm{c_r}^{(t2)})\Phi_{D}^\ast(\bm{d_r}^{(t2)})
 v_{\bar{a}\bar{b}}(|\bm{\bar{d}}-\bm{\bar{a}}|) \notag\\
 & \quad \times \Phi_{A}(\bm{a_r}^{(t2)})\Phi_{B}(\bm{b_r}^{(t2)}) ,
 \notag\\
 & \equiv (2\pi)^3 \delta^{(3)}(\bm{C}+\bm{D}-\bm{A}-\bm{B}) h^{(t2)}_{fi} ,
\end{align}
where 
\begin{align}
 \bm{a_r}^{(t2)} &
 = -\bm{\bar{a}} + \frac{\bar{m}_a}{m_a+\bar{m}_a}\bm{A} ,
 \\
 %%%
 \bm{b_r}^{(t2)} &
 = -\bm{\bar{d}} - \bm{C} + \frac{m_b}{m_b+\bar{m}_b}\bm{A} ,
 \\
 %%%
  \bm{c_r}^{(t2)} &
 = - \bm{\bar{a}} + \bm{A} - \frac{m_c}{m_c+\bar{m}_c}\bm{C} ,
 \\
 %%%
 \bm{d_r}^{(t2)} &
 = - \bm{\bar{d}} - \frac{\bar{m}_d}{m_d+\bar{m}_d}\bm{C} .
\end{align}

\section{\label{appendix:Fouriertrans}  Fourier transformations for deriving the local potential}

In this appendix, we perform the Fourier transformation to obtain the local potential in Eq.~\eqref{eq_localV}. In general, the integral part with any function $f^{(V)}(|k|)$ are given by
\begin{align}
 I^{(V)} =& \int\frac{d^3q}{(2\pi)^3} d^3p d^3k 
f^{(V)}(|k|) 
 e^{i\bm{q}\cdot\bm{r}} 
 \notag\\
&\times \exp\left[
 -\left(
 b_q \bm{q}\,^2 +  b_p \bm{p}\,^2  +  b_k \bm{k}\,^2 
\right.\right.
\notag\\
 &\left.\left.
    \quad\quad
 + 2 b_{qp}\bm{q}\cdot\bm{p}
 + 2 b_{pk}\bm{p}\cdot\bm{k}
 + 2 b_{kq}\bm{k}\cdot\bm{q}
 \right)
 \right]  
 \notag\\
 %%%
= & 
 \left(\frac{\pi}{b_p}\right)^{3/2}
 \int\frac{d^3 q}{(2\pi)^3}d^3k \, f^{(V)}(|k|) e^{i\bm{q}\cdot\bm{r}}
 \notag\\
& \times 
 \exp\left[
 -\left(
 c_q \bm{q}\,^2 + 2c_{qk} \bm{q} \cdot \bm{k} + c_k \bm{k}\,^2
 \right)
 \right] , 
\end{align}
where the coefficients are redefined by 
\begin{align}
&c_q = b_q - \frac{b_{qp}^2}{b_p} , \\
&c_{qk} = b_{kq}-\frac{b_{qp}b_{pk}}{b_p} , \\
&c_k = b_k - \frac{b_{pk}^2}{b_p} .
\end{align}

As a specific example, we demonstrate performing the integral for the capture 1.
In a similar way, the integrals for the other diagrams are also done by using the appropriate momenta shown in Appendix~\ref{appendix:BornAmp}.

In the case of the capture 1, 
the constants $b_q$, $b_p$, $b_k$, $b_{qp}$, $b_{pk}$, and $b_{kq}$ are given as follows.
\begin{align}
    b_q = & 
    -\frac{{b_A}^2 (R_a^m)^2}{8} -\frac{{b_B}^2 (R_b^m)^2}{8}-\frac{{b_C}^2 (R_c^m)^2}{8}-\frac{{b_D}^2 (R_d^m)^2}{8} , \\
    b_p = &
    -\frac{{b_A}^2}{2}-\frac{{b_B}^2}{2}-\frac{{b_C}^2}{2}-\frac{{b_D}^2}{2} , \\
    b_k = & 
        -\frac{{b_A}^2}{8}-\frac{{b_B}^2}{8}-\frac{{b_C}^2}{8}-\frac{{b_D}^2}{8} , \\
    b_{qp} = & 
        \frac{1}{2}\left(-\frac{{b_A}^2 {R_a^m}}{2}+\frac{{b_B}^2 {R_b^m}}{2}+\frac{{b_C}^2 {R_c^m}}{2}-\frac{{b_D}^2 {R_d^m}}{2}\right) , \\
    b_{pk} = &  
        \frac{1}{2}\left(\frac{{b_A}^2}{2}+\frac{{b_B}^2}{2}-\frac{{b_C}^2}{2}+\frac{{b_D}^2}{2}\right) , \\
    b_{kq} = &
        \frac{1}{2}\left(\frac{{b_A}^2 {R_a^m}}{4}-\frac{{b_B}^2 {R_b^m}}{4}+\frac{{b_C}^2 {R_c^m}}{4}+\frac{{b_D}^2 {R_d^m}}{4}\right) ,
\end{align}
where $R_i^m$ $(i=a,b,c,d)$ is given by
\begin{align}
    R_i^m = & \frac{m_i}{m_i+\bar{m}_i} .   
\end{align}

For hyperfine interaction with Gaussian potential $f^{(Hyp)}(|k|)= e^{-k^2/\Lambda^2}$, one obtain 
\begin{align}
 I^{(Hyp)} = & 
  \left(\frac{\pi}{b_p}\right)^{3/2}
 \int\frac{d^3 q}{(2\pi)^3}d^3k \,  e^{i\bm{q}\cdot\bm{r}}
    \notag \\
    & \times     
 \exp\left[
 -\left\{ 
 {\left(c_k + \frac{1}{\Lambda^2} \right)} \bm{k}\,^2
 + 2c_{qk} \bm{q} \cdot \bm{k}
 + c_q \bm{q}\,^2 
 \right\}
 \right] 
 \notag\\
= & \left(
 \frac{\pi}{4 b_p d_k e_q}
 \right)^{3/2}
 \exp\left[-\frac{r^2}{4e_q}\right] , 
\end{align}
where 
\begin{align}
    d_k = & c_k + \frac{1}{\Lambda^2} , \\
    e_q = & c_q - \frac{c_{qk}^2}{d_k} .
\end{align}

For the confinement potential 
$f^{(Conf)}(|k|) = \int d^3R\, R e^{-i\bm{k}\cdot\bm{R}}$, 
\begin{align}
 I^{(Conf)} = &
  \left(\frac{\pi}{b_p}\right)^{3/2}
 \int\frac{d^3 q}{(2\pi)^3}d^3k \,  e^{i\bm{q}\cdot\bm{r}}
 \int d^3R\, R e^{-i\bm{k}\cdot\bm{R}}
 \notag\\
 & \times
 \exp\left[
 -\left\{ 
 c_k  \bm{k}\,^2
 + 2c_{qk} \bm{q} \cdot \bm{k}
 + c_q \bm{q}\,^2 
 \right\}
 \right] 
 \notag\\
=  & 
 \frac{2\pi}{c_{rR}}
  \left(\frac{\pi}{4b_p c_k f_q}\right)^{3/2} \frac{1}{r}
 \exp\left(-\frac{r^2}{4 f_q} + \frac{c_{rR}^2}{4c_R}r^2\right)
 \notag\\
 & 
 \times 
 \left[\frac{3(d_r r)}{c_R}e^{-c_R (d_r r)^2} \right. 
 \notag\\
 & \quad
 \left. 
 -\sqrt{\frac{\pi}{c_R}}\left(\frac{1}{2c_R}+(d_r r)^2\right)
 {\rm erf}((d_r r)\sqrt{c_R})
 \right]  
\end{align}
where 
\begin{align}
&f_q = c_q - \frac{c_{qk}^2}{c_k} \, , \\
& c_{rR} = \frac{1}{2}\frac{c_{qk}}{f_q c_k} \, , \\
& c_R = \frac{1}{4}\left\{\frac{1}{c_k}+\frac{1}{f_q}\left(\frac{c_{qk}}{c_k}\right)^2\right\} \, ,  \\
& d_r = \frac{c_{rR}}{2c_R} \, . 
\end{align}

For the Coulomb potential, $f^{(Coul)}(|k|) = 1/k^2$, 
\begin{align}
 I^{(Coul)} = & 
 \left(\frac{\pi}{b_p}\right)^{3/2} 
 \int \frac{d^3 q}{(2\pi)^3} d^3k e^{i\bm{q}\cdot\bm{r}} \frac{1}{k^2}
 \notag\\
 & \times 
 \exp\left[
 -\left(c_q \bm{q}\,^2 + 2c_{qk}\bm{q}\cdot\bm{k}+c_k\bm{k}\,^2\right) 
 \right]
 \notag\\
=  & \frac{\pi^2}{4}\left(\frac{1}{b_p c_q}\right)^{3/2}
 \frac{c_q}{c_{qk}}\frac{1}{r}
 \exp\left(-\frac{r^2}{4C_q}\right) \mbox{erf}\left(\frac{c_{qk}}{2c_q \sqrt{g_k}}r\right),
\end{align}
where 
\begin{align}
    & g_k = c_k - \frac{c_{qk}^2}{c_q} \, . 
\end{align}

%
% ---- Bibliography ----
%

\bibliography{references}

\end{document}